**Topological insulator metamaterials**

Harish N. S. Krishnamoorthy, Alexander M. Dubrovkin, Giorgio Adamo, and Cesare Soci*

*Division of Physics and Applied Physics, School of Physical and Mathematical Sciences, Nanyang Technological University, 21 Nanyang Link, Singapore 637371*

*Centre for Disruptive Photonic Technologies, The Photonic Institute, Nanyang Technological University, 21 Nanyang Link, Singapore 637371*

Email: csoci@ntu.edu.sg

**Abstract**:

Confinement of electromagnetic fields at the subwavelength scale via metamaterial paradigms is an established method to engineer light-matter interaction in most common material systems, from insulators to semiconductors, from metals to superconductors. In recent years, this approach has been extended to the realm of topological materials, providing a new avenue to access nontrivial features of their electronic band structure. In this review, we survey various topological material classes from a photonics standpoint, including crystal growth and lithographic structuring methods. We discuss how exotic electronic features such as spin-selective Dirac plasmon polaritons in topological insulators or hyperbolic plasmon polaritons in Weyl semimetals may give rise to unconventional magneto-optic, non-linear and circular photogalvanic effects in metamaterials across the visible to infrared spectrum. Finally, we dwell on how these effects may be dynamically controlled by applying external perturbations in the form of electric and magnetic fields or ultrafast optical pulses. Through these examples and future perspectives, we argue that topological insulator, semimetal and superconductor metamaterials are unique systems to bridge the missing links between nanophotonic, electronic and spintronic technologies.

**Table of contents**

1. **Introduction**

Light-matter interaction underpins the functioning and further development of several photonic technologies. Metamaterial paradigms,[1,2] where materials are periodically structured at subwavelength scales, have been extensively exploited to achieve various regimes of coupling between highly confined electromagnetic fields and localized or propagating quasiparticles such as surface plasmon polaritons in metals and superconductors,[3] phonon polaritons in polar dielectrics,[4,5] and exciton polaritons in organic molecules and transition metal dichalcogenides.[6-8] In recent years, there has been growing interest in a class of materials exhibiting new quantum states of matter arising from topological features of their electronic band structure, namely topological insulators, semimetals and superconductors. These materials are characterized by the presence of gapless edge or surface states defined by the nontrivial topology of the bulk wavefunctions in the Hilbert space.[9,10] In the context of metamaterials, some of their unique properties are:[11,12]

i. The metallic surface states are topologically protected by time-reversal symmetry and cannot be destroyed or gapped by scattering processes, reducing plasmonic losses;[13,14]

ii. The spin of surface-state carriers is locked to the momentum so that spin states acquire helical polarization. Thus, the corresponding "spin-plasmons" can be controlled by the helicity of light;[15,16]

iii. Topological insulator crystals, particularly those of the chalcogenide family, have extremely high refractive index,[17] allowing strong optical confinement even in the dielectric regime.

iv. The inherent coupling of bulk dielectric states to metallic surface states enables tunability by charge injection, external electric and magnetic fields, or light polarization.[18,19]

Empowered by nanostructuring, topological materials provide a unique platform to combine nanophotonics with electronics and spintronics. Figure 1 shows a general classification for metamaterials based on topological systems. Depending on the dimensionality of the topological material and the presence of edge or surface states, metamaterials comprising of periodic arrangement of subwavelength metamolecules could couple plasmons of different nature:1D Dirac plasmons supported by the edge states of 2D topological insulators, 2D Dirac plasmons supported by the surface states of 3D topological insulators or 3D Dirac plasmons supported by the 2D surface states and connected by the Fermi-arc of a 3D Weyl semimetal.

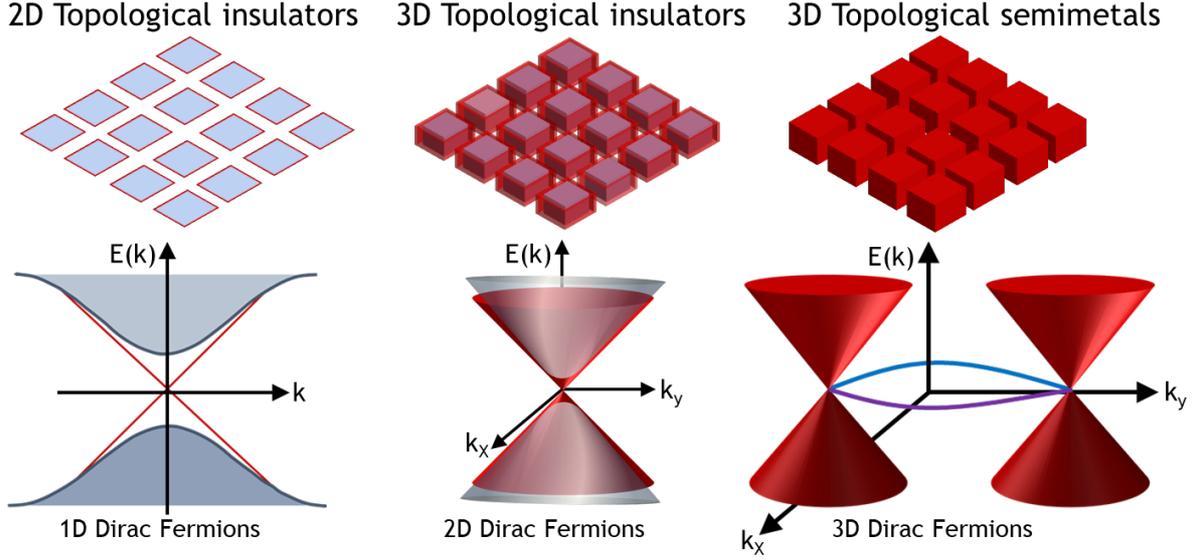

*Figure 1. Topological insulator and topological semimetal metamaterials with different dimensionalities: Classification of electromagnetic metamaterial concepts derived from different types of topological materials with their band dispersions. Such materials can be artificially structured at subwavelength scales to form metamaterials with unusual excitations. 2D materials feature 1D Dirac fermions at the edge of the metamolecules while 3D metamaterials feature 2D Dirac fermions at the surface of a topological insulator or 3D Dirac fermions traversing the bulk of a topological semimetal.*

This article aims to provide an overview of topological material systems suitable for the realization of artificially structured metamaterials and their functionalities. We start by discussing the general optoelectronic properties of 2D and 3D topological insulators, topological semimetals and superconductors with emphasis on the key optical signatures associated with topological states, which manifest in the dispersion of the optical response functions. We then focus our attention on chalcogenide topological insulator crystals which have been most commonly employed in metamaterial demonstrations. We discuss possible methods to grow and structure such crystals down to the subwavelength scale, including nanoparticle self-assembly and lithographic patterning of exfoliated crystals and thin films. We further survey plasmonic and dielectric properties of topological insulator nanostructures and existing demonstrations of photonic metamaterials realized in topological insulators and Weyl semimetals across the visible to the infrared spectrum. We finally provide some perspectives on possible development of artificially structured topological materials for hyperbolic and chiral topological plasmonics, non-Hermitian photonics, and their applications. We would like to point out that there are excellent reviews covering specific aspects of topological insulator materials such as their growth techniques[20,21] and electronic properties,[22-25] as well as Dirac plasmonics at THz frequencies.[26-28] With this article, we intend to provide an account of *structured* topological insulator systems across the electromagnetic spectrum, together with underpinning electromagnetic properties and growth aspects. We also dwell into emerging material systems such as topological semimetals and superconductors. Note that this review does not intend to provide a comprehensive description of the fields of metamaterials and topological insulators for which the reader is referred to the excellent reviews already available.[1,2,9,10] Furthermore, this review does not pertain to the field of topological photonics where suitably designed structures in topologically trivial materials lead to localization and transport of light via edge or surface states. The field of topological photonics has been extensively covered in other review articles.[29-31]

## 2. Topological materials for metamaterials

We will now discuss various topological material systems that can be employed for the realization of metamaterials including established platforms such as topological insulators and emerging platforms such as semimetals and superconductors, paying special attention to their optical properties.

### 2.1. Topological Insulators

The exotic features of topological materials arise from the bulk-boundary correspondence principle applicable to a system comprising an interface of a topologically nontrivial insulator with a trivial one. This is because the nontrivial insulator is characterized by a topological invariant which cannot change at the interface unless the bulk energy gap is closed, leading to the appearance of states within the gap.[10] Historically, topological properties of the band structure were first identified by Thouless, Kohmoto, Nightingale, and den Nijs (TKNN) in the integer quantum Hall (QH) effect exhibited by a 2D semiconductor under high magnetic field.[32] They showed that the topology of the nontrivial Hilbert space can be specified by an integer topological invariant called TKNN invariant $\nu$ such that the Hall conductivity $\sigma_{xy}$ is given by $\nu$ times $e^2/h$. $\nu$ is also called the first Chern number or the winding number and is equal to the Berry phase of the Bloch wave function calculated around the Brillouin zone (BZ) boundary divided by $2\pi$. Subsequently, the quantum spin Hall (QSH) effect was proposed. Unlike the QH effect, the QSH effect requires no magnetic field and preserves time reversal symmetry. QSH insulators are essentially two copies of the QH system, in which the chiral edge state is spin polarized and the two states form a time-reversed pair to recover the overall TRS, as shown schematically in Figure 2a.[33] Kane and Mele proposed a graphene model with spin-orbit coupling (SOC) to realize such a system.[34] They showed that a finite SOC leads to the opening of a gap at the crossing point of the Dirac cone (the Dirac point). Furthermore, within some parameter range, a time-reversed pair of spin-polarized one-dimensional (1D) states appears at the edges of the finite lattice. In this model, the desired spin polarization of the edge state is induced by the SOC which has an inherent tendency to align spins in relation to the momentum direction, generally referred to as *spin-momentum locking*. The description of topological invariants was extended to 3D systems by Moore and Balents who predicted the appearance of topologically protected states with Dirac cone dispersion.[35] In band insulators, the strong SOC leads to band inversion with the p-orbital valence band being pushed above the s-orbital conduction band, an effect commonly used to screen for new topological insulator materials. 2D and 3D topological insulator materials offer opportunities to couple electromagnetic waves with the topologically protected edge or surface states to create edge or surface plasmons, respectively. While practical realizations of topological insulator metamaterials will be discussed in Section 4, here we review the general optical properties of topological material systems, highlighting features related to surface states.

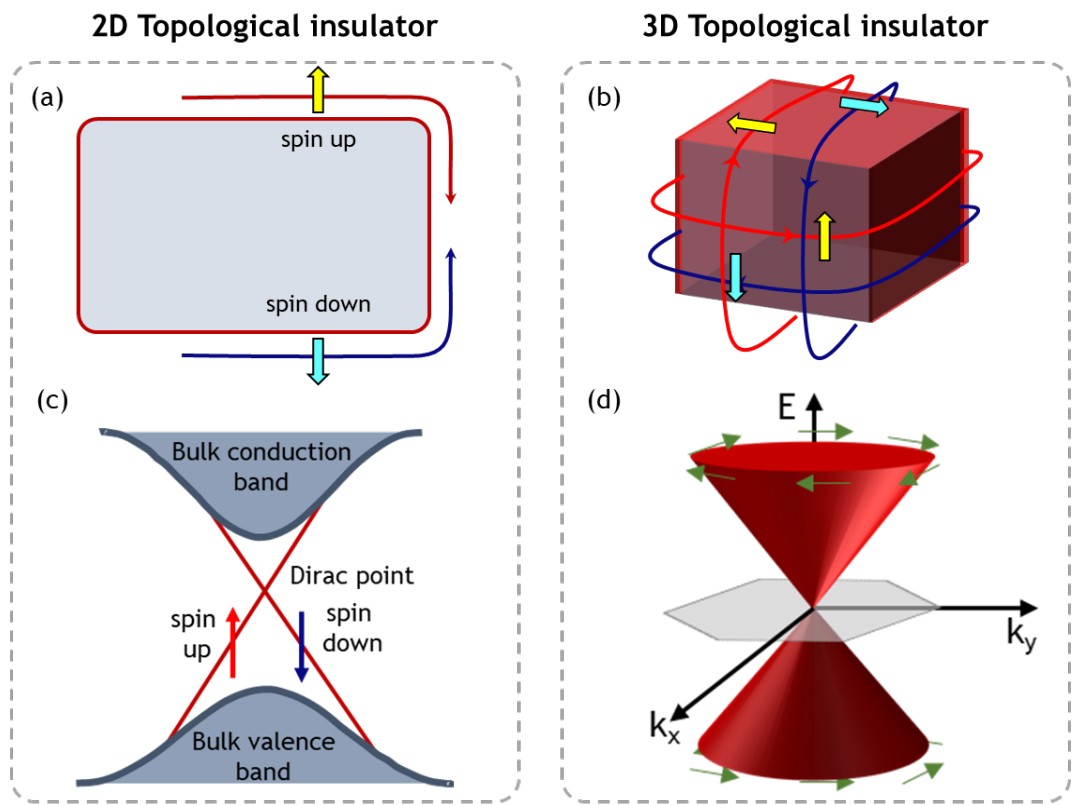

*Figure 2. Bulk-boundary correspondence and surface state dispersion:* Schematic showing (a) 2D (b) 3D topological insulator material systems with topologically protected, spin-polarized edge and surface states, respectively. These states arise from the non-trivial band structure of the bulk, which necessitates the appearance of topological surface states with Dirac dispersions at the interface with a trivial insulator or vacuum. Energy dispersion of (c) 2D edge states and (d) 3D surface states forming 1D and 2D Dirac cones, respectively.

### 2.1.1. 2D topological insulators

2D materials such as quantum wells, graphene-like Xenes and metal-organic frameworks are prototypical systems for the discovery and understanding of topological effects. Despite being the first systems where topological features were predicted theoretically and later demonstrated via QSH effects, realization of materials showing tangible optical response from topological edge states remains an ongoing effort.

*Quantum well structures*

The first 2D topological insulator to be experimentally demonstrated was a quantum well structure of CdTe/HgTe/CdTe wherein the presence of heavy metals Hg and Te leads to strong SOC effects and band inversion above a certain critical thickness (~ 6.3 nm) of the HgTe film.[36,37] In experiments, the QSH effect was manifested as a plateau in the residual conductance (~$2e^2$/h) of the quantum well structure. Similar effects were subsequently predicted and demonstrated in other quantum well systems such as AlSb/InAs/GaSb/AlSb.[38,39] Optically, the spin-polarized edge states of such quantum wells lead to polarization dependent properties such as the circular photogalvanic effect (refer to Section 2.1.2).[40]

*Graphene-like Xenes*

Graphene is a single layer of $sp^2$ hybridized carbon atoms laid out in a honeycomb lattice arrangement. Electrons in graphene are massless, characterized by a Dirac dispersion[41] and are the underlying factor for graphene's exceptional optical, electrical, thermal, and mechanical properties, promising a broad range of applications.[42,43] Graphene is a particularly attractive material for photonics owing to its ability for (i) supporting collective plasmonic excitations arising from electromagnetic excitations of free carriers as shown in Figure 3a and, (ii) exhibiting large change in optical properties upon application of a gate voltage that affects its strong interband transitions as shown in Figure 3b.[43-45]

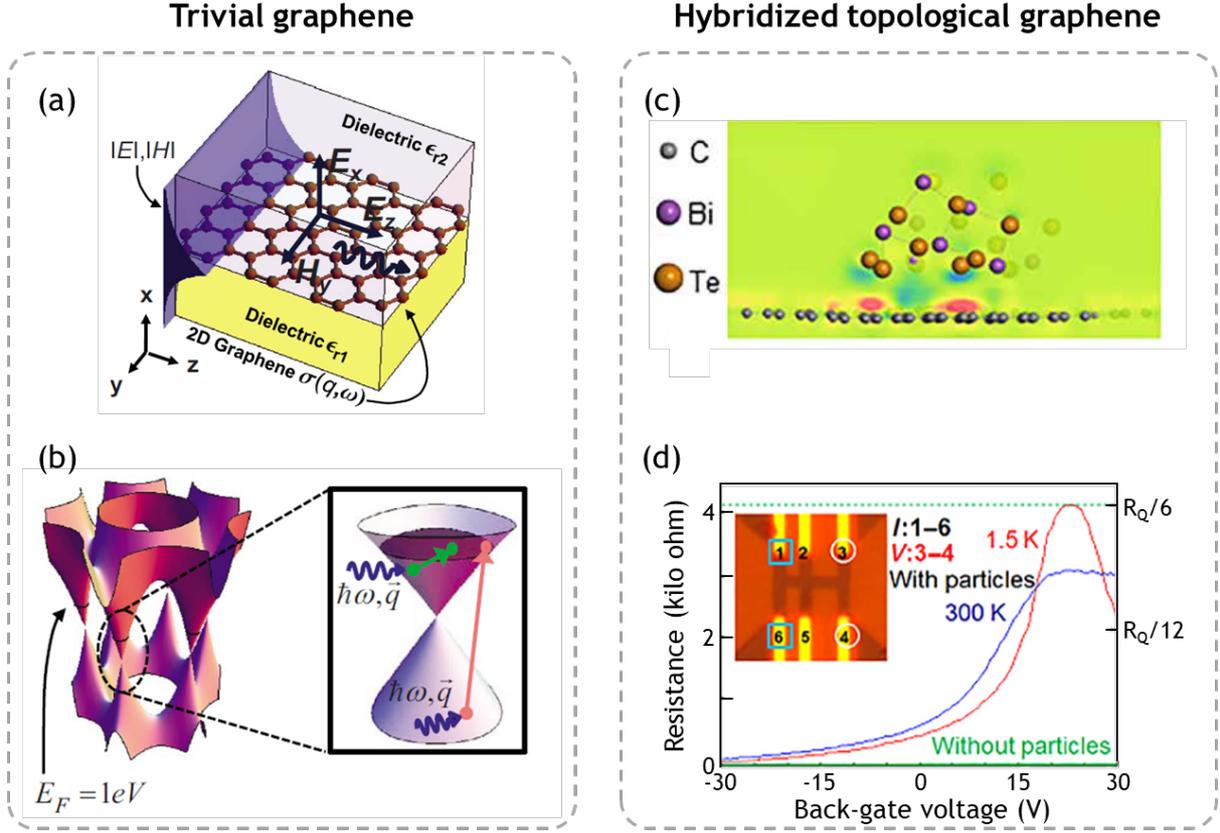

*Figure 3. Trivial graphene for plasmonics and demonstration of QSH graphene:* Schematic of a graphene photonic system composed of topologically trivial graphene layer sandwiched between two dielectrics, supporting TM plasmon polariton modes. (b) Electronic band structure of graphene, showing possible intraband (green) and interband (transitions) which contribute to absorption. (c)Atomic structure and charge-density difference of $Bi_2Te_3$ nanoparticles/graphene system. Red and blue colours indicate charge depletion and accumulation, respectively. (d) Four-probe resistance vs back-gate voltage measured on samples with (red/blue curves) and without (solid green curve) $Bi_2Te_3$ nanoparticles. In the former case, the resistance value approaches a rational fraction of the resistance quantum $R_Q$ indicative of edge transport. Figures a, b and c, d are reproduced with permission from Jablan et al.,[46] and Hatsuda et al.,[47] respectively.

While graphene was predicted to show QSH effect, experimental observation remained elusive due to its extremely small bulk bandgap ($10^{-3}$ meV) owing to a weak SOC. Since then, attempts have been made to increase the bulk gap and SOC. One prominent approach involves doping graphene with adatoms. For example, SOC enhancement has been reported in graphene with iridium nanocluster decoration[48], intercalation of Pb layer between graphene sheet and a Pt(111) surface has been shown to induce a large gap of about 200 meV with spin-resolved measurements confirming the spin-orbit nature of the bandgap.[49] While heavy adatom decoration is an effective approach to enhance the nontrivial bandgap and the QSH state in graphene, it also creates some undesirable consequences such as considerable shifting of the Dirac cone from the Fermi level.[50,51] An alternative approach that avoids such strong hybridization is to utilize systems with weak van der Waals interactions, which can enhance the SOC and induce the QSH state in graphene through interface interactions. Such proximity-enhanced QSH effect has been demonstrated experimentally in a system of graphene on top of few monolayers of WS$_2$ which has been shown to acquire SOC-induced bandgap of 17 meV,[52,53] approximately three orders of magnitude higher than the intrinsic value. Heterostructures in which graphene is sandwiched between layers of chalcogenides such as Sb$_2$Te$_3$ and MoTe$_2$ have been theoretically predicted to have large SOC bandgaps.[54,55] Hatsuda et al. demonstrated enhanced SOC in graphene decorated with dilute Bi2Te3 nanoparticles via four-probe resistance measurements where the measured resistance value approached a rational fraction of the resistance quantum R$_Q$ indicative of edge transport (Figure 3c, d).[47]

Apart from QSH graphene, a number of similar Xenes namely silicene, germanene, stanene, plumbene and bismuthene are expected to exhibit a topological gap and related optoelectronic properties. Of these, large band gaps have been experimentally demonstrated in bismuthene[56] and stanene.[57] A survey of such materials as potential THz metamaterials has been put forward in a recent review by Lupi and Molle.[26]

*Organic topological insulators*

Strong SOC, which is the key requirement for producing nontrivial bandgaps, can also be realized at the molecular level. There is an ongoing quest for discovering new classes of topological insulator systems beyond conventional two-dimensional quantum well systems. QSH states have been predicted in organic materials consisting of covalent organic frameworks (COFs) composed of heavy metal atoms like Pb and Bi self-assembled into an hexagonal lattice.[58] This new class of organic topological insulators was predicted to show nontrivial topological edge states that are robust against significant lattice strain. Other organometallic framework compounds, for example, Ni$_3$C$_{12}$S$_{12}$,[59] 2D indium-phenylene organometallic framework (IPOF),[60] and Cu−dicyanoanthracene (DCA) lattice,[59] were also predicted to exhibit topologically nontrivial behaviour. While many of these compounds have been successfully synthesized, their nontrivial character is yet to be demonstrated.

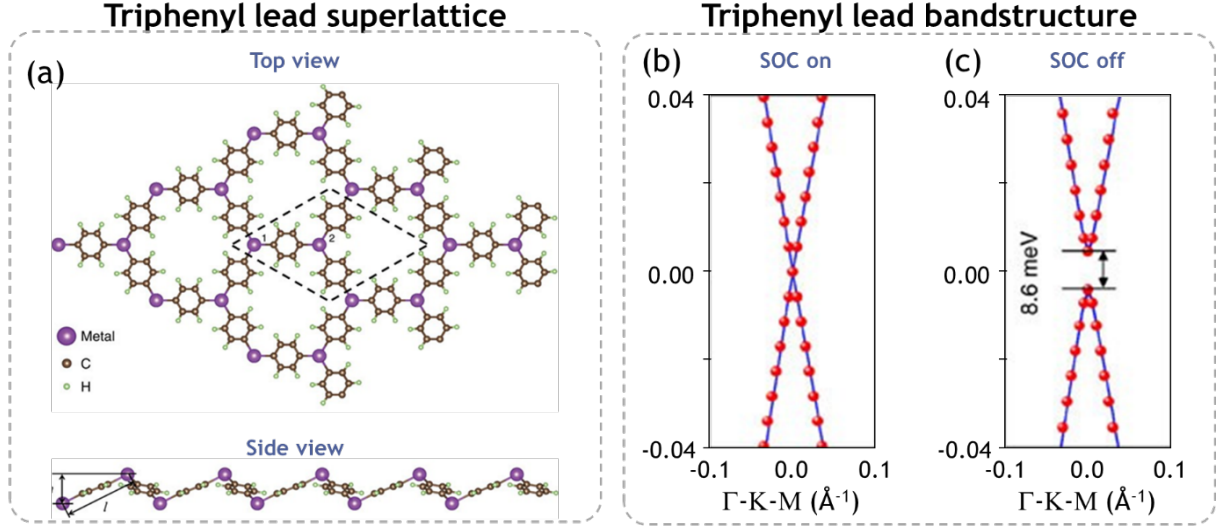

*Figure 4. Organic topological insulators:* (a) Top and side view of the 2D organometallic triphenyl lead superlattice. (b) and (c) Band structures of triphenyl-lead (TL) lattice without and with spin-orbit coupling (SOC), respectively. Blue lines are obtained from DFT calculations while red dots are the fitted bands using the effective Hamiltonian. Figures reproduced with permission from Wang et. al.[58]

### 2.1.2. 3D topological insulators

Unlike 2D topological materials, the following features make 3D topological materials suitable for photonic applications across the electromagnetic spectrum: (i) absorption from bulk phonon modes in the THz, (ii) strong bulk polarizability and high refractive index in the infrared, (iii) bulk interband absorption and negative dielectric constant in the visible region, and (iv) Drude response from both 2D Dirac surface fermions and 3D massive bulk fermions across the infrared/THz regions.

The first material to be predicted and experimentally identified to be a 3D topological insulator was $Bi_xSb_{1-x}$.[61,62] However, its complicated surface state structure with multiple Fermi level crossings made it difficult to characterize. A whole family of chalcogenides with tetradymite structure belonging to the $Bi_xSb_{2-x}Te_ySe_{3-y}$ (BSTS) family was later identified as a class of 3D topological insulators featuring a simpler surface state structure. Among these materials, experimental evidence of Dirac surface states in $Bi_2Se_3$, $Bi_2Te_3$ and $Sb_2Te_3$ were shown in 2009.[63] The two materials, $Bi_2Se_3$ and $Bi_2Te_3$ lie at the opposite ends of the compositional spectrum with the largest (~0.3 eV) and smallest (~0.17 eV) bandgaps, respectively. The three-dimensional bi-chalcogenide compounds have a rhombohedral crystallographic structure arranged in the order Bi(Sb)-Se(Te)-Bi(Sb)-Se(Te), with all of them belonging to the space group $R\bar{3}m$. As shown in Figure 5a, they comprise of a layered structure with quintuple layer (QL) unit cell, wherein the interlayer interaction is mediated by van der Waals forces.[13] Band structures of the materials were determined by first principles approaches using density functional theory (DFT) as shown in Figure 5b for three representative compositions.[13,64,65] In such materials, the strong SOC is brought about by the presence of heavy elements such as Bi and Sb. Inclusion of SOC in the DFT calculations immediately manifests in the form of band inversion of the bulk band structure.

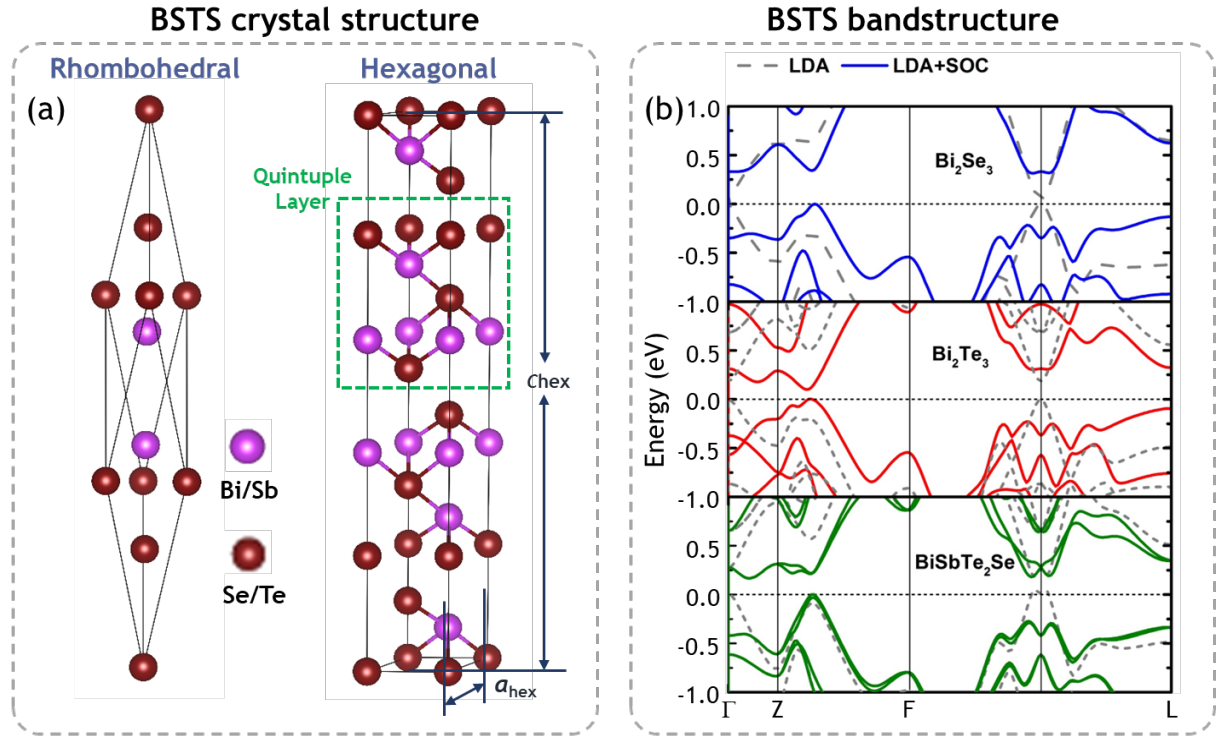

*Figure 5. Crystal structure and band structure of $Bi_xSb_{2-x}Te_ySe_{3-y}$ (BSTS) topological insulators:* (a) Crystal structure of BSTS compounds exhibiting topological insulator behaviour - rhombohedral and equivalent hexagonal lattice cell, with a quintuple layer (QL) shown by the green rectangle. (b) Band structure of representative binary and quaternary $Bi_xSb_{2-x}Te_ySe_{3-y}$ compounds with and without spin-orbit coupling retrieved from density functional theory (DFT) calculations. Figures a and b are respectively reproduced and adapted with permission from Jun et al.[13]

*Bulk optical properties* – The optical response of bulk BSTS shows some characteristic spectral features: the low frequency THz and far-infrared response (well below gap) features the high polarizability of the bulk together with absorption from vibrational modes (for example, $Bi_2Se_3$ shows α and β phonon modes at 1.85 THz and 4.0 THz respectively) as seen in Figure 6.[66,67] Depending on the intrinsic doping, free carriers in the bulk may also contribute to the optical response.[68,69] The mid-infrared region below the bulk bandgap is also characterized by a high polarizability, giving rise to extremely high refractive index values. This is an attractive feature for dielectric nanophotonic architectures, particularly in the infrared spectral region where these chalcogenide materials have very low optical losses.[13,17,67]

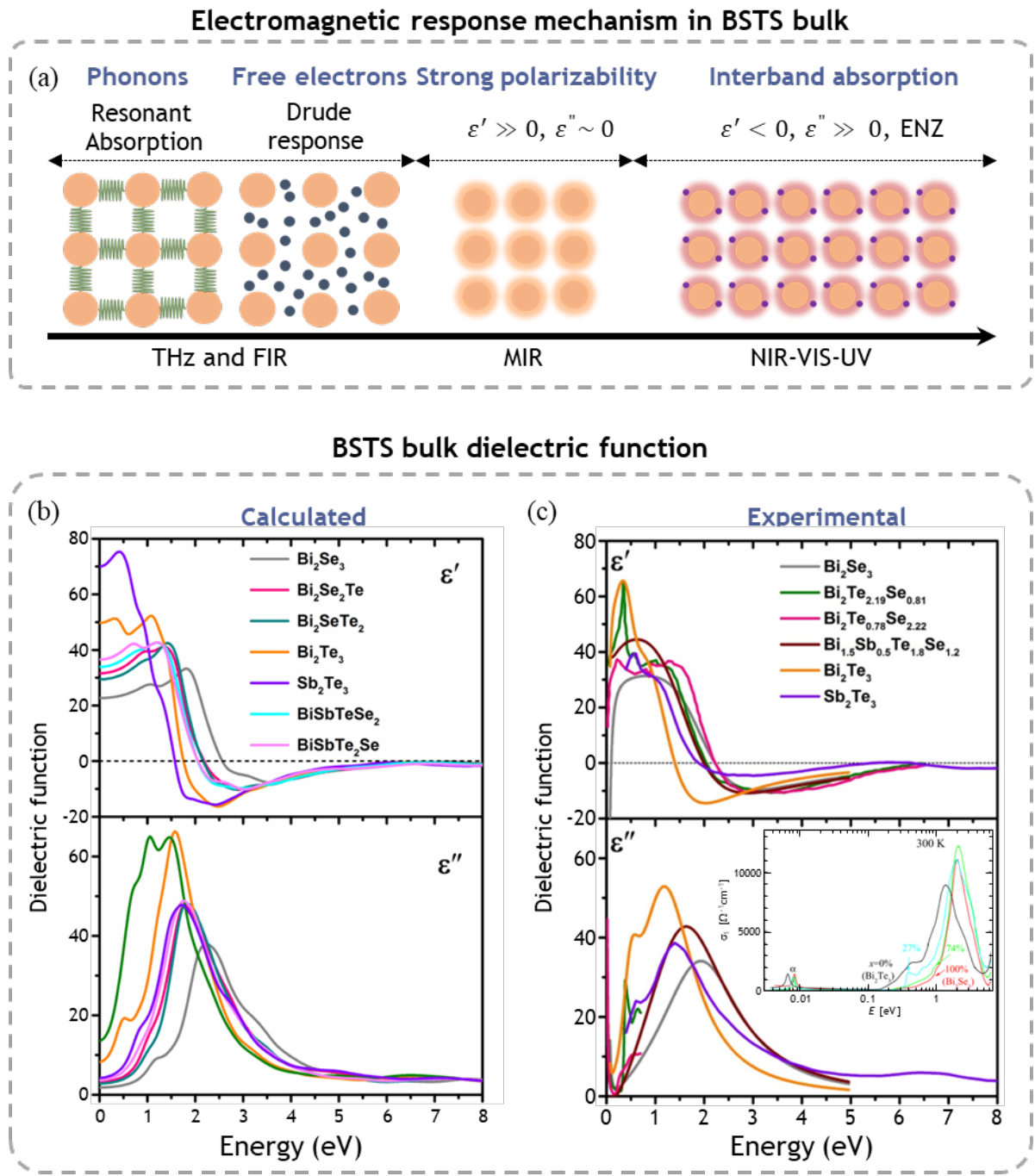

*Figure 6. Optical response of the bulk:* (a) Schematic showing the electromagnetic response features arising from the bulk across the spectrum. Real ($\varepsilon'$) and imaginary parts ($\varepsilon''$) of the bulk dielectric function for various compositions of chalcogenide $Bi_xSb_{2-x}Te_ySe_{3-y}$ topological insulator compounds based on (b) DFT calculations, and (c) experimental measurements. Inset of (c) shows the optical conductivity of $Bi_2(Te_{1-x}Se_x)_3$ depicting the emergence of the $\alpha$ phonon at low frequencies. Figures b and c are reproduced and adapted with permission from Jun et al.[13] Inset of Figure c is reproduced with permission from Dubroka et al.[67]

Figure 6 shows calculated and experimentally measured complex dielectric functions ($\tilde{\varepsilon} = \varepsilon' + i\varepsilon''$) of several BSTS materials.[13] The calculated functions are derived by DFT while the experimental functions are determined using a combination of optical spectroscopy and ellipsometry.[17,67] The materials are characterized by a negative $\varepsilon'$ in the near-infrared to ultraviolet frequency region accompanied by large $\varepsilon''$ arising from interband absorption. In the infrared (energy < 1 eV), the materials feature a high polarizability (large $\varepsilon'$) and low losses (small $\varepsilon''$), while in the visible and ultraviolet regions, the materials show plasmonic response ($\varepsilon' < 0$) with higher losses (large $\varepsilon''$). The optical dielectric function of the bulk is typically modelled by employing a Tauc-Lorentz oscillator to model the interband absorption along with Lorentz oscillators for phonon absorption and Drude oscillator to account for response from free bulk carriers.

*Surface state plasmonic properties* – Strong SOC in BSTS materials results in the formation of surface states exhibiting a Dirac dispersion with typical carrier concentrations of the order of $10^{13}$ cm$^{-2}$. Calculations based on first principles can be employed to estimate the carrier concentrations from which the 2D DC, AC conductivities and in turn, the dielectric function can be determined.[13] An example is shown in Figure 7 where the band structure of a 5 quintuple layer $Bi_2Se_3$ slab shows Dirac dispersion associated with the surface bands as well as the surface charge distributions. Calculations show that BSTS materials exhibit strong, low-loss plasmonic behaviour in the low frequency regions (mid-infrared to THz) as shown in Figure 7c. Despite the presence of marked Dirac surface states, the greatest challenge for the realization of plasmonic metamaterials is the reduction of intrinsic doping. For example, in $Bi_2Te_3$, the Dirac point is located beneath the top of the valence band, which makes it difficult to probe the surface response without being disturbed by bulk carriers. Generally, binary 3D topological insulators, including $Bi_2Se_3$ and $Sb_2Te_3$, are characterized by high bulk carrier concentrations, particularly in samples obtained from bulk single crystals.[68,70,71]

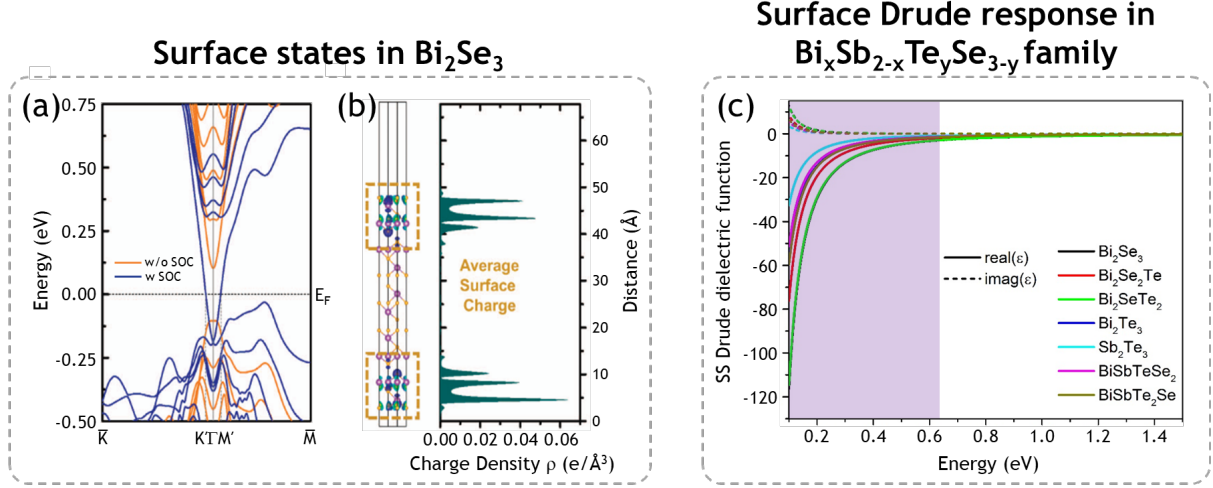

*Figure 7. Band structure of ultrathin TI layer and surface Drude response*: (a) Band structure of 5QL $Bi_2Se_3$ slab without and with inclusion of spin-orbit coupling effects (the horizontal dashed line indicates the Fermi energy level and K'-M' the region of occupied surface states). (b) Three- and one-dimensional charge density distribution of conducting surface states; all occupied orbitals below the Fermi level and their degeneracy were considered to calculate the average charge density. (c) Real and imaginary parts of the dielectric function arising from 2D Drude response of the surface states. Figures (a) and (b) reproduced with permission from Jun et. al.,[13] and (c) adapted from Jun et al. [13]

Conversely, ternary and quaternary BSTS compounds of the form $Bi_xSb_{2-x}Te_ySe_{3-y}$, wherein a compensatory approach is used to combine n-type ($Bi_2Se_3$ and $Bi_2Te_3$) with p-type ($Sb_2Te_3$) alloys, have significantly lower bulk conductivity.[72-75] Specific compositions for which the optical properties are known include $Bi_2Se_2Te$, $Bi_2Te_2Se$,[76] $(Bi,Sb)_2Te_3$,[19,77] $BiSbTeSe_2$,[78] $Bi_{1.5}Sb_{0.5}Te_{1.7}Se_{1.3}$,[78] and $Bi_{1.5}Sb_{0.5}Te_{1.8}Se_{1.2}$.[79] Reijnders et. al. showed signatures of surface states at low temperatures in $Bi_2Te_2Se$ using a combination of low temperature Fourier Transform Infrared (FTIR) optical spectroscopy and ellipsometry measurements,[76] suggesting that phonon-mediated coupling between bulk and surface states suppresses surface conductance as temperature rises. By employing terahertz time-domain spectroscopy (THz-TDS), Tang et. al. showed that in a highly compensated quaternary topological insulator, $Bi_{1.5}Sb_{0.5}Te_{1.8}Se_{1.2}$, the experimentally retrieved Drude spectral weight approached theoretical predictions.[79]

Another feature of these van der Waals materials is that they exhibit strong anisotropy with hyperbolic dispersion over certain parts of the electromagnetic spectrum, exhibiting a similar behavior as artificially engineered layered hyperbolic metamaterials.[80-85] This is a consequence of the quintuple-layered structure as well as the resulting strong anisotropy in the bulk phonon modes. In particular, propagation of subdiffractional hyperbolic phonon-polariton modes influenced by the Dirac plasmons arising from the topological surface states in thin flakes of $Bi_2Se_3$ and $Bi_2Te_{2.2}Se_{0.8}$ have been shown by near-field nanoscopy.[82] Esslinger et. al., extracted the dielectric tensor components of $Bi_2Se_3$ and $Bi_2Te_3$ and showed hyperbolic dispersion in the near-infrared to visible part of the spectrum.[80] Lingstädt and co-workers exploited the same property in $Bi_2Se_3$ to show propagation of polaritons around defects using cathodoluminescence spectroscopy and electron energy-loss spectroscopy.[81]

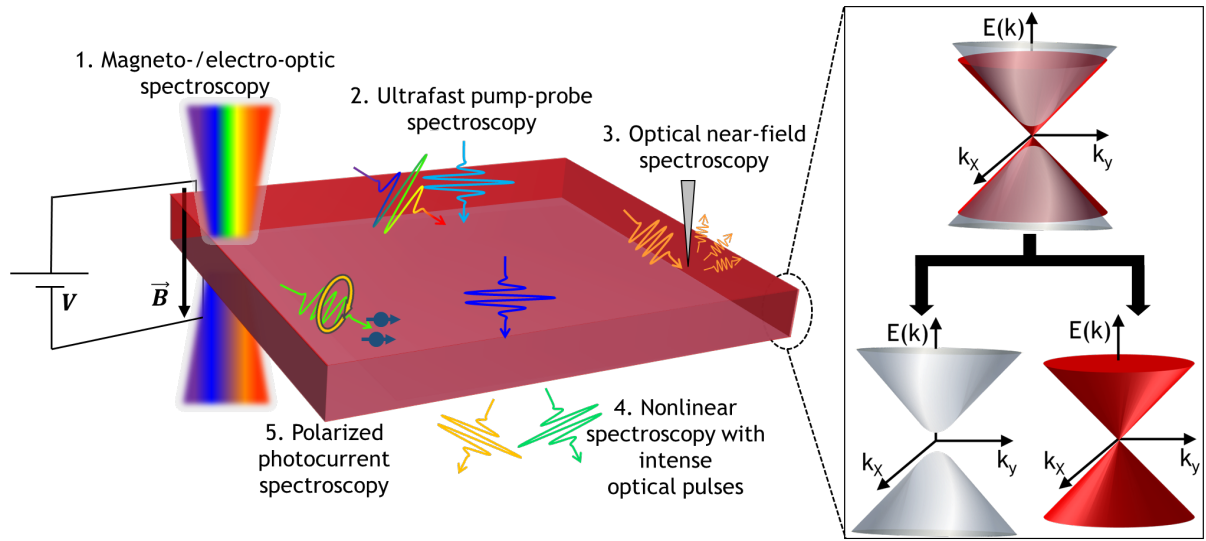

*Figure 8. Techniques for decoupling surface from bulk response: Schematic showing various optical approaches employed to decouple responses of bulk and surface states. Far-field spectroscopic measurements under external magnetic field/bias (1), or with ultrafast pump and probe beams (2) can give rise to a measurable change in the optical response from the surface state. Near-field optical spectroscopy (3) also facilitates isolation of the surface state response by measuring evanescent waves. Use of intense optical pulses enables higher harmonic generation from the surface states (4). Driving the population of spin-polarized Dirac carriers out of equilibrium by circularly polarized light excitation generates electrical currents (5).*

*Decoupling surface from bulk response* – The main challenge to isolate surface contributions from the dominant bulk response in topological insulator materials can be addressed using surface selective techniques. Besides the broadly employed methods for the determination of electronic properties of topological insulator, such as electron photoemission spectroscopy[62,86,87] or scanning-tunnelling microscopy[88-91], surface-sensitive electro-/magneto-optical and all-optical spectroscopic techniques were successfully employed to study topological insulator materials (Figure 8).

1. Magneto-/ Electro-optic spectroscopy: Application of large external magnetic fields affects the spin orbit coupling, and the corresponding variation in optical response can be employed to discriminate surface from bulk contributions.[68,92,93] For example, Faraday rotation was used to differentiate electron and hole responses via the sign of the Faraday angle, while Landau level transitions as a function of magnetic field showed distinct linear and parabolic dependencies for bulk and Dirac surface carriers, respectively (Figures 9a, b).[19] Yet another approach employing magnetic field relied on the notion that topological insulators are best described as bulk magnetoelectrics, with quantized magnetoelectric coefficient set by the fine-structure constant $\alpha$.[94]

Far-field spectroscopy measurements under an external bias were used to elicit and control responses from bulk and surface state carriers. Whitney et al. showed that the spectral weight between bulk and topological surface channels, as well as between interband and intraband channels, can be varied by an applied gate voltage in a film of BST (Figure 9c).[18] This manifested as a change in the differential reflectance and transmission spectra induced by the competition between bulk-driven and surface-driven effects at different gate voltages.

2. Ultrafast pump-probe spectroscopy: Time-resolved terahertz spectroscopy is a useful tool to directly probe the collective response of low-energy electronic transitions.[95-97] Sim et al. employed ultrafast optical-pump terahertz-probe spectroscopy and showed nontrivial surface electron dynamics of hot Dirac electrons due to opening of a carrier relaxation channel from bulk to surface state.[95] The study also revealed characteristics such as surface state scattering rates to be ~ hundreds of fs, much longer than that of non-topological systems such as Drude metals. Panna et al. employed broadband time-resolved transient pump-probe reflectivity measurements at visible-near infrared frequencies with co-circularly polarized light to demonstrate efficient linear-optical access to ultrafast spin dynamics of surface states in $Bi_2Se_3$.[96]

3. Optical near-field spectroscopy: Surface sensitive techniques such as scanning near-field optical microscopy enable measurement of evanescent waves and thus facilitate direct measurement of plasmon modes associated with the topological surface state. Near-field spectroscopy was broadly applied to isolate the response of topological surface states.[14,82,98,99] Chen et al. showed real-space spectroscopic THz near-field images of thin $Bi_2Se_3$ layers revealing polaritons with amplitude decay time $\tau = 0.48$ ps, similar to the decay times measured by far-field extinction spectroscopy of $Bi_2Se_3$ ribbons (Figure 9d).[100] Pogna et al. used a combination of hyperspectral time domain spectroscopy, nano-imaging and detectorless scattering near-field optical microscopy to map propagation of collective modes in single crystal flakes of $Bi_2Se_3$ and $Bi_2Te_{2.2}Se_{0.8}$ topological insulators.[82]

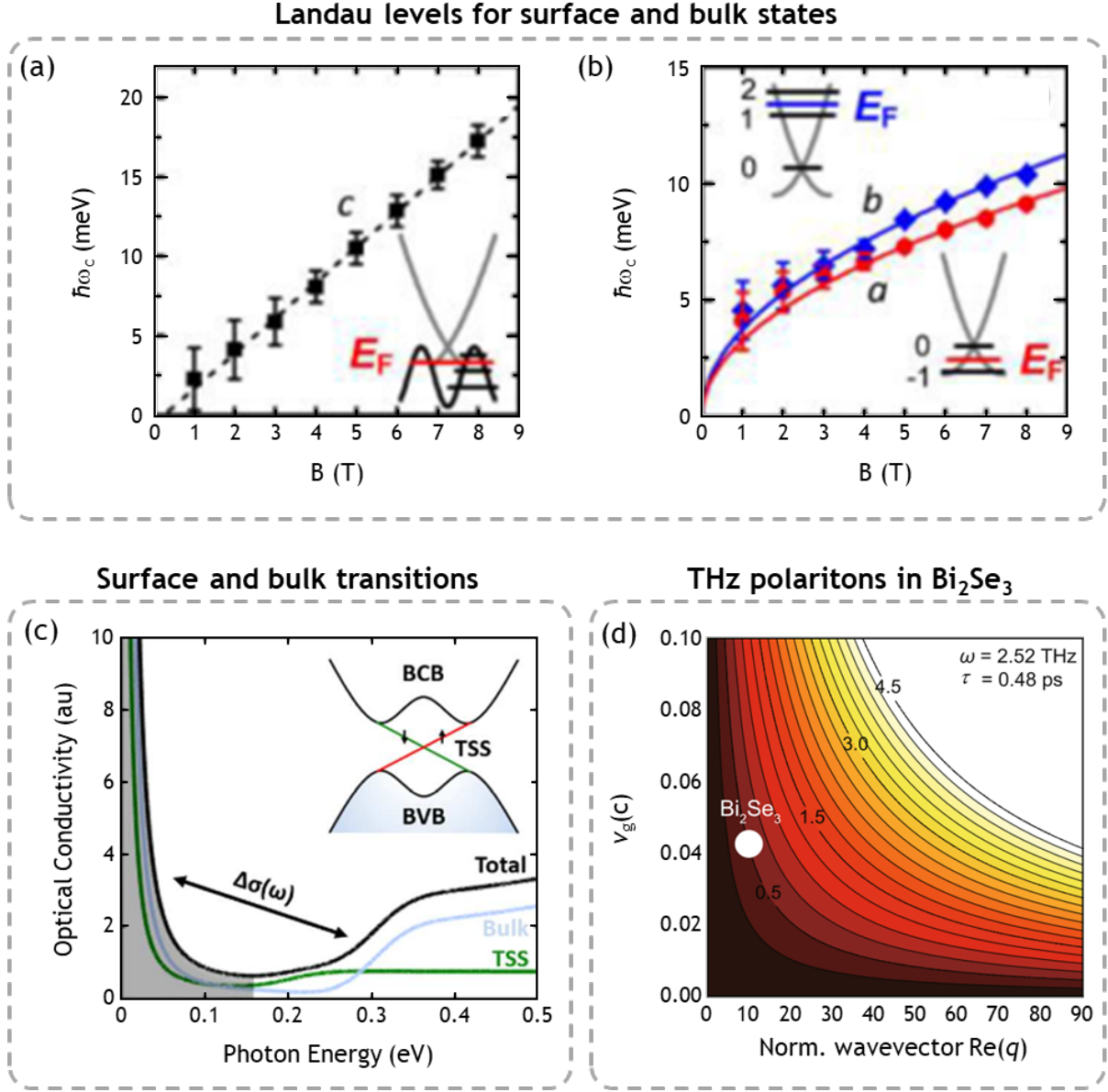

*Figure 9. Optical response induced by surface states in topological insulators:* *Far-field spectroscopic measurements under external magnetic field and the measurement of resulting Landau level transition gives markedly different responses from the bulk and surface carriers, with the former showing a linear dependence (a), and the latter showing a parabolic dependence on the applied magnetic field (b).[19] (c) Far-field spectroscopic measurements under external bias can drive optical transitions where the spectral weight is transferred between both bulk (blue) and topological surface channels (green).[18] (d) Relative propagation length $L/\lambda_p$ of 2.52 THz polaritons as a function of wavevector and group velocity for $Bi_2Se_3$.[14] $\tau = 0.48$ ps is the polariton amplitude decay time. Numbers in black are the relative propagation lengths. White symbol corresponds to experimental relative propagation length for THz polaritons in the 25-nm-thick $Bi_2Se_3$ determined from scanning near-field spectroscopic measurements. Figures a and b are reproduced with permission from Shao et al.,[19] while c and d are reproduced with permission from Whitney et al.,[18] and Chen et al.,[14] respectively.*

4. Nonlinear spectroscopy: McIver et al. postulated and demonstrated second harmonic generation (SHG) of light from the surface of the topological insulator $Bi_2Se_3$.[101] By measuring the intensity and polarization of the emitted second harmonic frequency light as a function of crystal orientation, they determined the relevant susceptibility tensor elements that contribute to SHG and also showed that SHG intensity depends on the surface carrier concentration, and thus can be employed to optically probe changes in the surface Fermi level. Giorgianni et al. showed strong nonlinear optical response from Dirac carriers arising from harmonics of the intraband Dirac current under an oscillating electromagnetic field.[102] Intensity dependence of transmission scaling as $(E_0/E_{max})^6$ at large fields was attributed to third harmonic generation from surface states with efficiency of ~ 1%, comparable to that of graphene. Recently, Xiang et al. showed third harmonic generation with a nonlinear conversion efficiency of 0.01% in chalcogenide topological insulators $Bi_2Se_3$, $Bi_2Te_3$ and $Sb_2Te_3$ and attributed it to the nontrivial topological ordering of the bands.[103]

5. Polarization dependent photocurrent spectroscopy: The optoelectronic response of topological materials under circularly polarized illumination shows peculiar effects. When the topological medium is driven out of equilibrium through circularly polarized optical excitation, the balance of surface carriers with opposite spin is altered and spin currents are transformed into spin-momentum–locked net electrical currents. This behaviour can be phenomenologically represented by the equation[104,105]:

$$j(\varphi) = Csin2\varphi + L_1 sin4\varphi + L_2 cos4\varphi + D$$

The coefficients D and L2 are arise out of photocurrent contributions from the semiconducting bulk. Specifically, D is related to the polarization-independent photothermoelectric background current, which sets the overall directional current flow, while L2 to the photon drag effect, which results from linear momentum transfer of the incident photons to the excited carriers. In contrast, the coefficients L1 and C quantify photocurrent contributions from surface carriers. Such currents, driven by linear and circular polarization of the incident light, originate from the linear and circular photogalvanic effects, respectively.

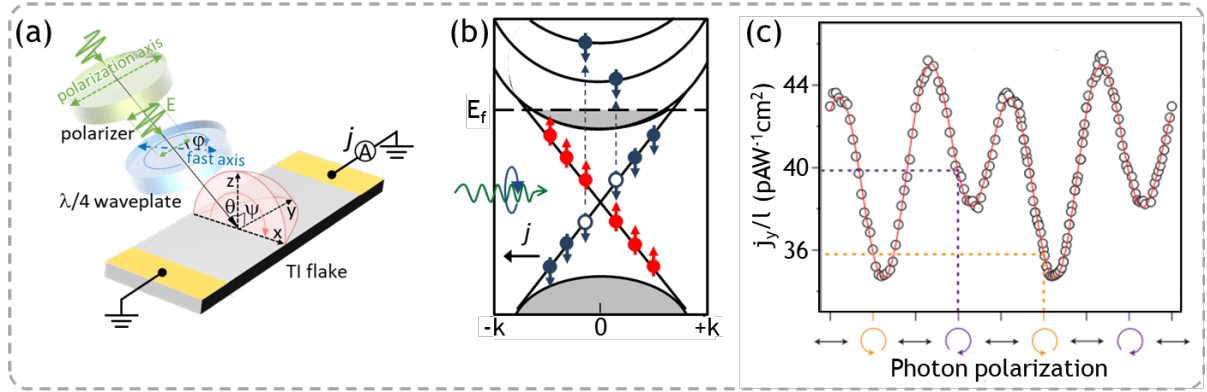

*Figure 10. Polarization dependent photocurrent:* (a) Experimental scheme for studying helicity dependent photocurrent in topological insulators. (b) Excitation by circularly polarized light drives the population of the spin-polarized carriers out of equilibrium leading to current flow. (c) Experimentally measured photocurrent in an exfoliated flake of $Bi_2Se_3$ as a function of incident photon polarization. Figures (a), (b) and (c) are reproduced with permission from Sun et al.,[16] and McIver et al.,[104] respectively.

Hosur theoretically predicted the generation of helicity-dependent direct currents by circularly polarized light excitation of the surface states of a topological insulator.[106] The first instance of circular photogalvanic effect in a 2D topological insulator system was shown by Wittman et. al., using HgTe/CdHgTe quantum well structures.[40] McIver et al. then showed the same in a 3D topological insulator system – $Bi_2Se_3$ topological insulator flakes when illuminated with circularly polarized light, generated a photocurrent originating from topological Dirac fermions, and its direction could be reversed by reversing the helicity of the incident light.[104] In the same work the authors also identified a component of the photocurrent, of topological surface state origin, which depended on the linear polarization of the incident light. Notwithstanding the fact that helicity dependent photocurrents (HDPC) can unravel the spin nature of surface electrons in topological insulators, their detection is hindered by large non-topological currents generated by the semiconducting bulk, induced by unintentional doping. Various strategies have been devised to magnify the surface contribution to the HDPC, such as using electrical gating to tune the Fermi energy,[107] exciting the surface carriers below the bulk bandgap,[108] thinning the topological insulator flakes down a few quintuple layers,[109] or grow chalcogenide topological insulator crystals with low intrinsic doping.[75]

## 2.2. Emerging material platforms

### 2.2.1. Topological semimetals

Semimetals are materials characterized by a small or vanishing density of states at the Fermi level. Topological semimetals are a particular group of semimetals in which the finite density of states results from the crossing of valence and conduction bands in the 3D Brillouin zone characterized by a topological invariant.[110] Unlike topological insulators, topological semimetals support 3D fermions and are classified depending on the dimensionality of band crossings (Figure 11). The most studied group is that of so-called nodal semimetals with 0D band crossings (nodal points) – sub-categorized into Dirac semimetals[111] and Weyl semimetals,[112] and 1D band crossings along lines in the momentum space, called nodal-line semimetals.[113] These intriguing topological phases can form when two bands with distinct-conserved quantum numbers interpenetrate, forming a semimetallic nodal ring of degenerate states with Dirac dispersion or further shrink down to discrete points (Figure 11a).[114]

The key interest in these materials lies in the unconventional optical and transport properties which stem from their linear band dispersion. Unlike topological insulators, the dispersion of the optical response in topological semimetals is characterized by peculiar features in the interband optical conductivity at different frequency regions due to the quasi-linear band crossings. Generally, the interband optical response of systems characterized by electron–hole symmetric d-dimensional ungapped linear band crossings yields a universal power-law frequency dependence of the real part of the conductivity, $\sigma_1(\omega) \propto \omega^{d-2}$.[115] For the case of three-dimensional topological semimetals (d=3), this gives $\sigma_1 \propto \omega$, that is a linear frequency dependence. In recent years, this fundamental feature of topological semimetal materials has been verified experimentally by optical spectroscopy and has been found to show different behaviour depending on the type of topological semimetal. For example, the Dirac semimetal $Cd_3As_2$[116] shows nearly perfect linear $\sigma_1(\omega)$ in the mid-infrared, weakly dispersive nodal-line semimetal ZrSiS[117] exhibits a nearly flat band behaviour of the conductivity spectra and Weyl semimetals such as TaAs shows linear in frequency spectra with different slopes at different regions of the spectrum.[112] Such rich behaviour in conductivity may be exploited to create plasmonic responses. For example, Hue et al. reported plasmonic standing waves in nanoribbons of the Dirac semimetal $PtTe_2$ in the midinfrared range.[118] The observed plasmonic response disappeared when the material thickness reduced below 10 nm, which was ascribed to the intrinsic layer-dependent optoelectronic properties of the topological semimetal crystal.

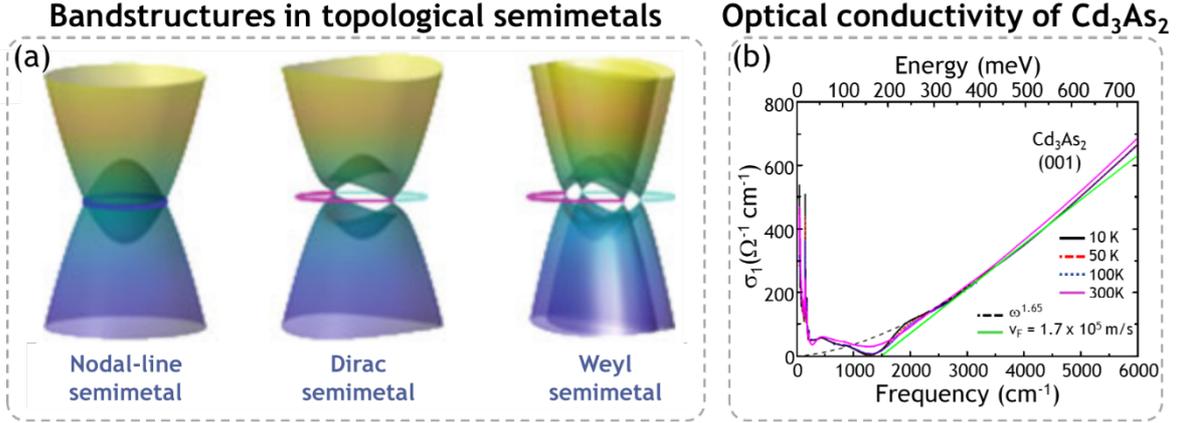

*Figure 11. Optical response of topological semimetals:* (a) Schematics of the band structures in topological semimetals.,[114] (b) Real part of the optical conductivity in Dirac semimetal $Cd_3As_2$ measured at various temperatures.[116] Figures a and b are reproduced with permission from Beidenkopf et al.,[114] and Neubauer et al., respectively.[116]

Topological semimetals are also promising platforms for nonlinear photonics. A family of transition metal monopnictide Weyl semimetals TaAs, TaP and NbAs has been shown to yield giant second harmonic generation.[119,120] In Dirac semimetals such as $Cd_3As_2$, efficient higher harmonic generation at THz frequencies has been attributed to a combination of coherent acceleration of Dirac electrons in momentum space, nonlinear kinetics of electron distribution as well as the linear Dirac dispersion.[121,122] In addition, other aspects such as nonlinear photoresponse as well as saturable absorption in such materials have been employed for applications such as photodetection and mode-locked lasing.[123-125]

In addition, topological semimetals are theoretically predicted to support rich electrodynamics resulting in unconventional propagation of light.[126-134] This includes tunable response to external perturbations such as heat or magnetic fields, as well as hyperbolic dispersion. For example, $Cd_3As_2$ shows thermo-optic shifts larger than conventional III-V semiconductors[135] while $Co_3Sn_2S_2$ exhibits giant magneto-optical response. Low-energy optical transition related to the Weyl point/anti-crossing line has been suggested to be the main mechanism behind the giant magneto-optical transitions.[136,137] Natural hyperbolic optical properties of a candidate Weyl semimetal $WeTe_2$ enabled to realize far-infrared 2D hyperbolic plasmons in the range between 16 to 23 μm.[138] Recently, 3D hyperbolic plasmons in layered nodal-line semimetal ZrSiSe was demonstrated in the telecom band via near-field microscopy measurements.[139] Topological semimetals are also suitable candidates to generate highly chiral photocurrents. Very promising preliminary studies were conducted by Ma et al. who detected chirality of the Weyl fermions through the photocurrent generated in TaAs crystals by circularly polarized mid-infrared light illumination,[140] and by Ji et al. who investigated the spatially dispersive nature of the circular photogalvanic effect in $Mo_xW_{1-x}Te_2$ semimetal.[141]

### 2.2.2. Topological superconductors

Topological superconductors are materials exhibiting a trivial spin-degenerate superconducting gap in the bulk and a topologically nontrivial spin-polarized superconducting gap at the interfaces (Figure 12a). Contrary to a topological insulator whose surface state features Dirac electrons, a topological superconductor's surface state is composed of Majorana fermions, particles that coincide with their own antiparticles. Topological superconductors are characterized by features such as:[23,142] (i) odd parity pairing symmetry with a full superconducting gap; (ii) gapless surface states composed of Majorana fermions; and (iii) Majorana zero modes in the superconducting vortex cores. Material candidates showing topological superconductivity are still emerging. So far, experimental signatures have been observed in hybrid systems, such as a 2D magnetic material or a topological insulator with a superconductor,[143-147] and a few intrinsic materials, such as Fe (Te, Se) and $UTe_2$.[148-151] Topological superconductors are yet to be explored experimentally from an optical perspective. However, there have been theoretical studies and proposals on their potential as platforms for optics and photonics.[152-155]

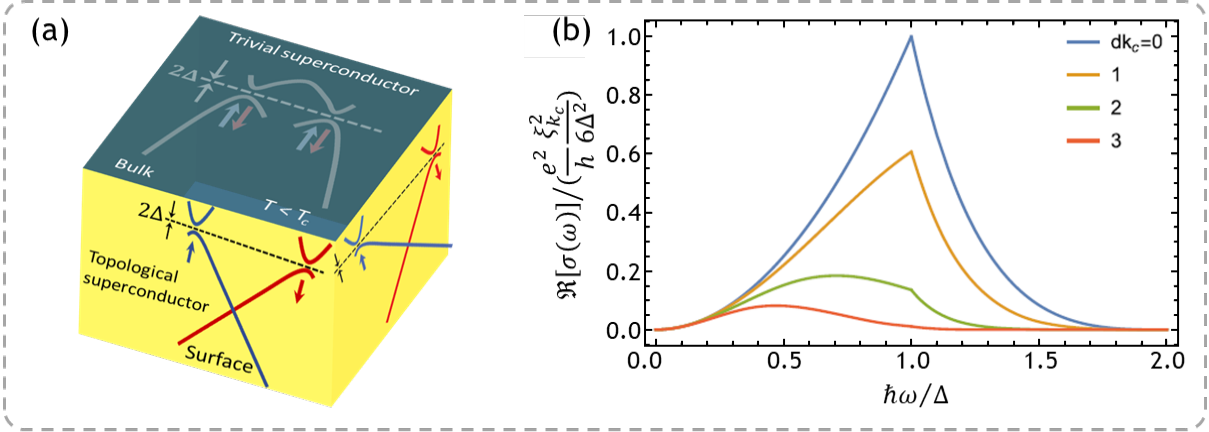

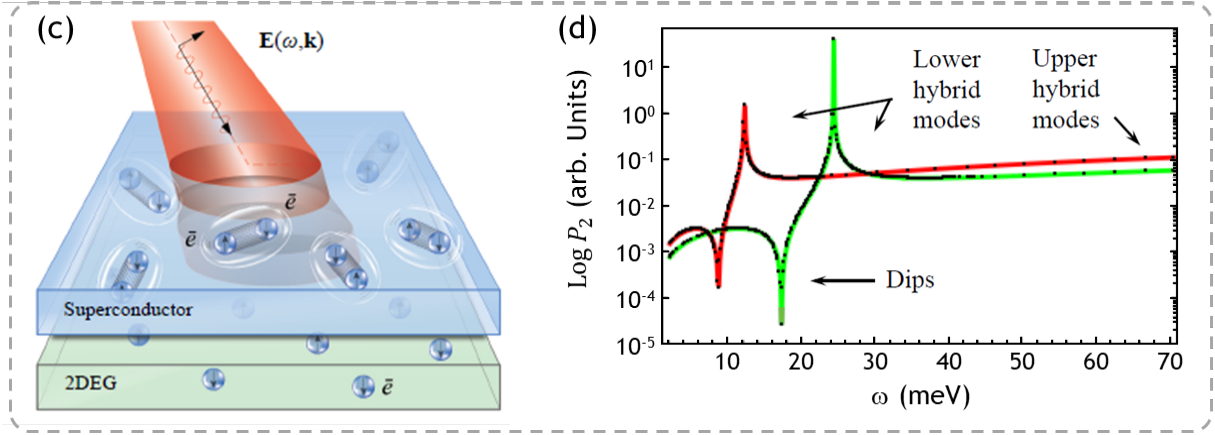

*Figure 12. Topological superconductors for photonics: (a) Schematic of the band structure of a 3D (2D) topological superconductor characterized by trivially superconducting spin-degenerate bulk states and topologically superconducting spin-polarized surface (edge) states. (b) Calculated real part of optical conductivity due to Majorana states of a topological superconductor as a result of absorption of photons breaking Cooper pairs into Majorana modes.[153] (c) Schematic of a hybrid normal metal–superconductor structure exposed to an electromagnetic field of incident light.[155] (d) Power absorption spectrum as a function of in-plane wavevector for such a system in topologically non-trivial case showing hybrid modes created by intralayer, interlayer and light-matter interactions. Figures b, and c, d are reproduced with permission from He et al.,[153] and Villegas et al., respectively.[155]*

Claassen et al. showed that the out-of-equilibrium superconducting state in chiral topological superconductors can be controlled on ultrafast timescales.[156] Yu et al. proposed the implementation of a Hadamard gate using the Majorana modes of optically engineered domains in a chiral topological superconductor where short optical pulses could be used to write, erase and move chiral domains.[154] He et al. calculated the optical conductivity of a topological superconductor due to absorption of photons considering that the ground state with chiral Majorana modes consists of a Cooper pair condensate and the absorption of photons breaks them into Majorana modes (Figure 12b).[153] They showed that Majorana fermions have relatively strong optical signature which can be detected by measuring the local optical conductivity. Villegas et al. proposed another optical approach based on a hybrid system of superconductor coupled to a material with plasmonic gapless excitations such as a topological insulator (Figure 12c). Their calculations showed that the interaction of light with the superconductor is highly amplified in such a system, manifesting itself as giant Fano-like resonances that uniquely characterize the Majorana fermions (Figure 12d).[155]

3. From topological insulator materials to metamaterials

Topological insulators are synthesized using a variety of techniques which may be categorized based on the dimensionality of the resulting crystals. Besides bulk crystals, topological materials are also extensively studied in the form of thin films and nanostructured crystals (such as nanoplates and nanoribbons) with large surface to volume ratio, where surface properties become dominant. In such systems, nanolithographic methods, such as electron beam and focused ion beam lithography, have been recently employed to create artificial structures that couple to and enhance the topological surface state response. This section provides a general overview on the growth of chalcogenide topological insulator materials and the methods to structure or assemble them into metamaterials.

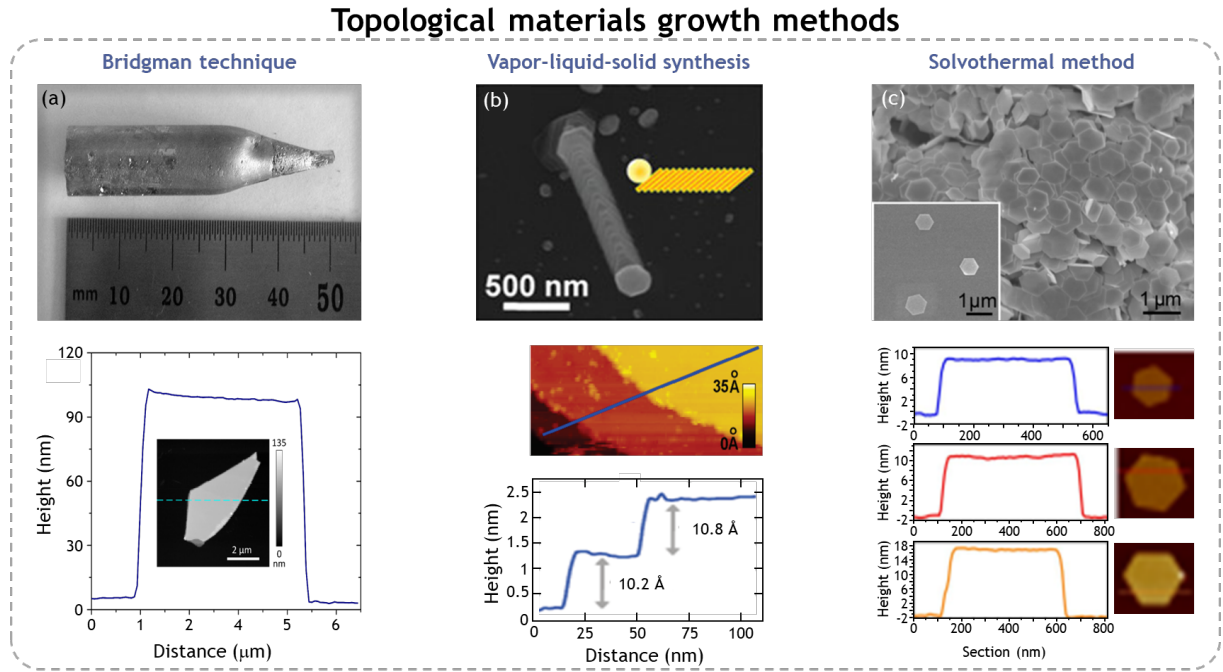

*Figure 13. Different forms of topological materials :* (a) Photograph of a crystal of $Bi_2Te_3$ grown by Bridgman technique (top panel).[157] Atomic force microscopy (AFM) scan profile and image (inset) of a flake of $Bi_{1.5}Sb_{0.5}Te_{1.8}Se_{1.2}$ exfoliated from bulk crystal. (b) SEM image of a nanowire grown off c-axis using Au-catalysed vapor-liquid-solid (VLS) synthesis (top panel).[158] AFM image showing terraces on the nanoribbon surface (middle panel) and line-cut showing the ribbon height across three steps (bottom panel) (c) Scanning electron microscopy (SEM) image showing cluster and individual nanoplates synthesized by solvothermal method (top panel).[159]

*Topographical AFM image of three representative nanoplates of thicknesses 9 nm, 12 nm and 18 nm (bottom panels). Top panel of Figure a is reproduced with permission from Atuchin et al.[157] Figures b and c are reproduced with permission from Kong et al.,[158] and Zhao et al., respectively.[159]*

### 3.1. Synthesis methods of topological insulators
### 3.1.1. Growth of topological bulk crystals

The Bridgman–Stockbarger technique or the Bridgman method is a common technique employed for the growth of bulk single crystals from the liquid phase. In this approach, a vertical crucible containing the melt is moved slowly relative to a temperature gradient, nucleation begins at the tip of the sample as it enters the cooler zone and growth from stable domains continues with the growth front moving into the melt. Crystals several centimetres long have been grown successfully using this technique.[157,160-162] For example, Atuchin et al. showed the growth of high quality single crystals of $Bi_2Te_3$ employing an ampoule translation rate of 10 mm/day and an axial temperature gradient of approximately 15 °C/cm (Figure 13a).[157]

Another technique in practice is that of flux growth, wherein the necessary components of the melt are first dissolved into a solvent called the flux. The relative concentrations of the flux and solids of interest are typically decided based on thermodynamic and technical considerations such as melting temperatures, reactivity of melt, and volatility. The flux can also be one or more of the elements in the targeted compound, in which case the growth is referred to as a self-flux process. For example, Sultana et. al., grew samples of single crystal $Bi_2Te_3$ by self-flux from high temperature (950 °C) melt and slow cooling (2 °C/hour).[163] Single crystals of other compositions such as $Sb_2Te_3$, $Bi_2TeI$ have also been grown.[164,165]

Apart from these, vapor phase techniques such as physical vapor transport (sublimation of the solid material followed by condensation)[166-168] or chemical vapor transport (involving chemical reactions between the starting materials and the transport agent)[169,170] have also been employed to produce bulk crystals of topological materials. For example, Atuchin et al. showed the growth of $Bi_2Te_3$ microcrystals with dimensions of approximately 100 μm by physical vapor transport.[167] Jiao et al. demonstrated the growth of single crystals of $Bi_2Se_3$ with dimensions as large as 3×3×1 $mm^3$ using iodine as the transport agent.[169]

For further details on growth techniques of bulk topological insulator crystals, we refer the reader to the comprehensive review paper by May et. al.[20]

### 3.1.2. Thin film deposition

*Mechanical exfoliation* – Adopted from the graphene community,[41,171] where thin flakes are cleaved from bulk crystals by an adhesive tape, this method is applicable to materials with layered structures such as chalcogenide topological insulators where the adjacent layers are held together by weak van der Waals forces. It is worth noting that, unlike graphene, chalcogenide crystals are more fragile and susceptible to damage during the exfoliation process. Nonetheless, single crystals with thicknesses on the order of the unit cell can be isolated from bulk samples, yet with small lateral dimensions.[172] A typical AFM profile of an exfoliated flake of $Bi_{1.5}Sb_{0.5}Te_{1.8}Se_{1.2}$ is shown in Figure 13a.

*Molecular Beam Epitaxy (MBE)* – This is the most widely employed technique to deposit high-quality large area thin crystalline films with well-defined stoichiometry in ultra-high vacuum (UHV) environment ($\sim 10^{-9}$ torr). The essential mechanism is based on vapor phase transport where the elemental source materials are heated to high temperatures and the vapours deposit on the surface of a crystalline substrate. MBE allows the growth of films with finely controlled stoichiometry and doping levels, notably compensated ternary or quaternary topological insulator chalcogenide crystals with high bulk resistivity,[173-175] topological superconductors and heterostructures. For example, superconductor-topological insulator heterostructures of $Bi_2Se_3$ and $NbSe_2$ have shown superconducting order induced in the topological surface states through the proximity effect.[176,177] Another important feature of MBE for optical applications is that it allows growth of topological insulators on a variety of substrates such as silicon, sapphire, graphene, GaAs, and $SrTiO_3$.[176-179] The reader is referred to the paper by Ginley et al., for an extensive review of MBE growth of topological insulators.[21]

*Other methods* – Alternative chemical or physical deposition methods were successfully employed for the growth of thin films and nanostructured topological insulator crystals, including metal organic chemical vapor deposition (MOCVD)[180-182] and pulsed laser deposition (PLD).[183]

### 3.1.3. Synthesis of nanostructured crystals

*Chemical vapor deposition (CVD)* – Nanocrystals in the form of platelets and ribbons are commonly grown using methods such as vapour-liquid-solid (VLS) or vapour-solid (VS) techniques in a tube furnace. In the case of VLS, a catalyst such as Au nanoparticles helps in promoting uniaxial crystal growth. The VLS method involves three stages starting with metal alloying with the vapor, followed by nucleation of the crystal and, finally, axial growth of the crystal. For example, $Bi_2Se_3$ nanostructures have been commonly synthesized using this method, where the structural parameters of the nanostructures may be controlled by factors such as substrate temperature and gas flow rate (Figure 13b).[158,184] $Bi_2Se_3$ nanowires are formed by stacking of nanoplatelets. The nanowire growth direction is determined by the way of stacking, either parallel to c-axis or off c-axis. These nanowires usually range from several hundred nanometers to 20 µm in length, consisting of hundreds of nanoplatelets.[158]

*Solvothermal synthesis* – This is a widely adopted technique for the growth of nanostructures that involves precursors dissolved in organic solutions. The precursors react with each other under supercritical conditions forming nanostructures of the desired compound. The size and shape distribution as well as crystallinity of the nanostructures are controlled by the reaction temperature and time, solvent and precursors. High quality $A_2^{V}B_3^{VI}$-type alloy hexagonal nanoplatelets of various compositions have been synthesized based on modified reduction reaction between suitable oxides ($Bi_2O_3$, $Sb_2O_3$, $TeO_2$, $SeO_2$) and ethylene glycol (Figure 13c)[185] and shown to display topological properties.[159]

### 3.2. Lithographic methods for metamaterial fabrication

So far, artificial structuring of topological insulators for fabrication of metamaterials has been accomplished either using resist-based indirect patterning techniques such as electron beam lithography[100,186] and photolithography, or direct patterning by focused ion beam milling by highly energetic ions[17,187] (Figure 14).

Electron beam lithography that employs selective exposure, development and removal of a polymer resist and subsequent reactive ion etching of the patterned areas (Figure 14a) enables fabrication of structures with critical dimensions as small as 40-50 nm with smooth sidewalls (an example of a microring metamaterial structure fabricated using electron beam lithography is shown in Figure 14b).[186]

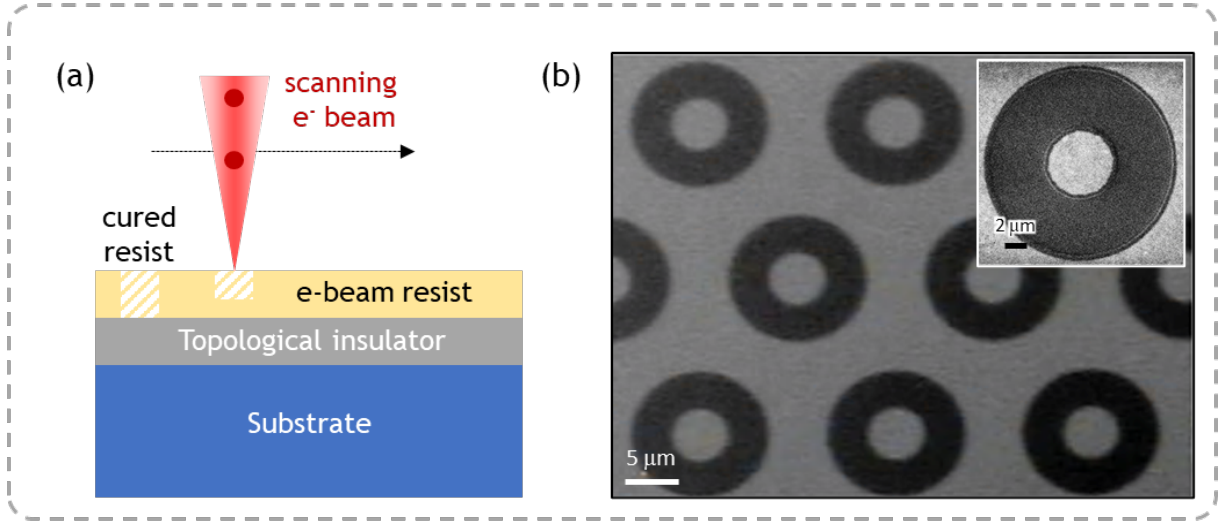

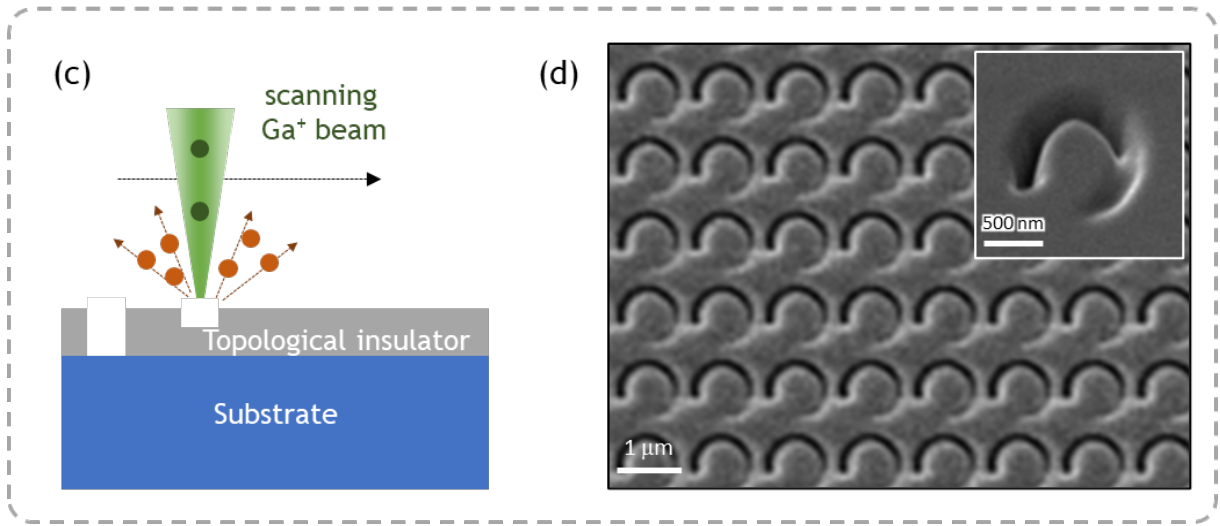

*Figure 14. **Techniques for fabrication of topological insulator metamaterials***: *(a) Indirect patterning by electron beam lithography, which uses a polymer resist sensitive to the electron beam. The desired pattern is defined on the resist by scanning the electron beam across the sample and is transferred to the underlying topological insulator film by reactive ion etching. (b) A microring metamaterial array fabricated by electron beam lithography. (c) Direct patterning of the topological insulator using a focussed beam of highly energetic gallium ions. (d) A 3D spiral metamaterial array fabricated by focussed ion beam milling. Figure b is reproduced with permission from Autore et al.[186]*

Most commonly, focused ion beam milling has allowed direct writing of nanoscale patterns on topological insulators upon rastering a beam of highly energetic, heavy ions from an appropriate source (typically Ga$^+$) across the sample (Figure 14c). The high-energy ions remove the material within the point of impact on the sample, enabling generation of metamaterial patterns with high precision and resolution down to few tens of nanometres. Focused ion beam milling is suitable for structuring samples with relatively small lateral dimensions and enables fabrication of complex 3D structures such as spiral metamaterials (Figure 14d).

Both techniques have their pros and cons. Electron beam lithography enables fabrication of structures with small critical sizes but generally requires a sample with larger lateral area and is unsuitable for working with exfoliated flakes which are typically quite small. In addition, one may need to consider undesired effects of the electron beam resist and etch gas on the topological properties of the sample. Focused ion beam lithography is highly suited to working with exfoliated flakes as it does not involve the spin coating of a resist but can sometimes result in the generation of residual debris of the sputtered material as well as undesirable ion implantation which may impact the topological character of the sample.

## 4. Tunable, active and reconfigurable topological insulator metamaterials

Strongly confined electromagnetic fields produced in artificially structured metamaterials can be employed to couple to the optical excitations of topological materials discussed in Section 3. For example, simple metamaterial structures such as arrays of gratings, slits and split ring resonators (Figure 15a) can be utilized to excite surface plasmon polariton modes of 2D Dirac carriers and manipulate their interaction with bulk phonon modes. Structures with polarization dependent response may also enable optical control of spin-polarized surface state carriers due to the inherent spin-momentum locking of topological insulators. Furthermore, when combined with the high refractive index of the bulk of topological insulator crystals, metamaterials could be used to induce dielectric multipole resonances and surface currents. Overall, the creation of subwavelength domains in topological insulator materials has enabled enhancement and engineering of several light-matter interaction phenomena such as topologically protected Dirac plasmons (Figure 16a),[188] hybridization of Dirac plasmons and phonons (Figure 16b),[100,186,189,190] magnetoplasmons,[191] highly confined electromagnetic fields resulting in complex mode structures and induced surface currents(Figure 16c),[17] enhanced photocurrents[16] and near-field response,[99] as well as structural colours.[187]

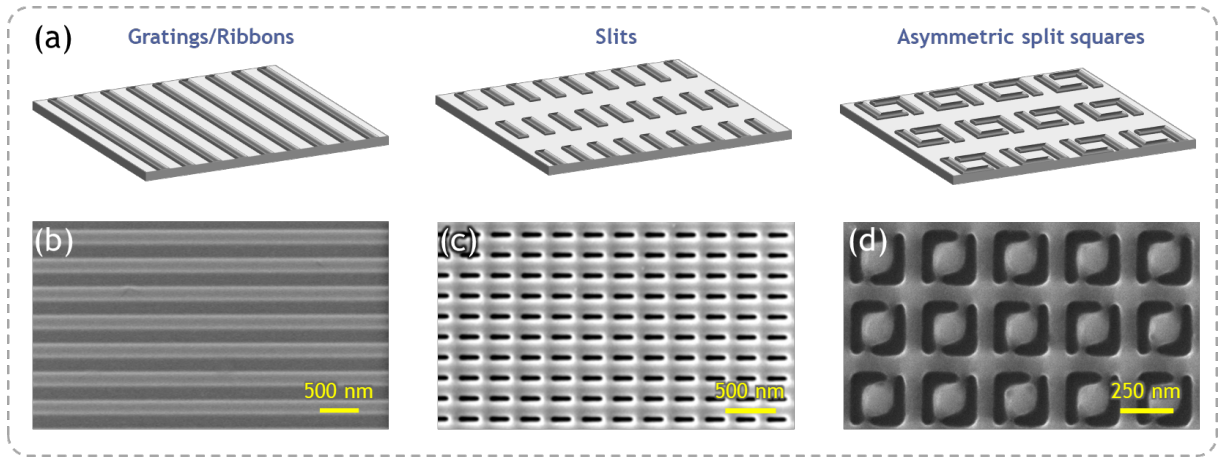

*Figure 15. **Schemes for topological insulator metamaterials**: (a) Schematic of some commonly employed metamaterial structures in topological materials, (from left to right) gratings or ribbons, slits and asymmetric split squares. (b), (c) and (d) SEM images of corresponding structures fabricated by nanolithographic techniques. Scale bars are 500 nm, 1 μm and 250 nm for (b), (c) and (d), respectively. Figure b is reproduced with permission from Ou. et. al.[187]*

In the sections to follow, we discuss the optical properties of topological insulator metamaterials across the electromagnetic spectrum, and their tunability from THz and infrared to visible and ultraviolet frequency regimes. We also give examples of active and reconfigurable topological insulator metamaterials with enhanced optical response to electric and magnetic fields, ultrafast optical pulses or circularly polarized light inducing spin-dependent circular photogalvanic currents.

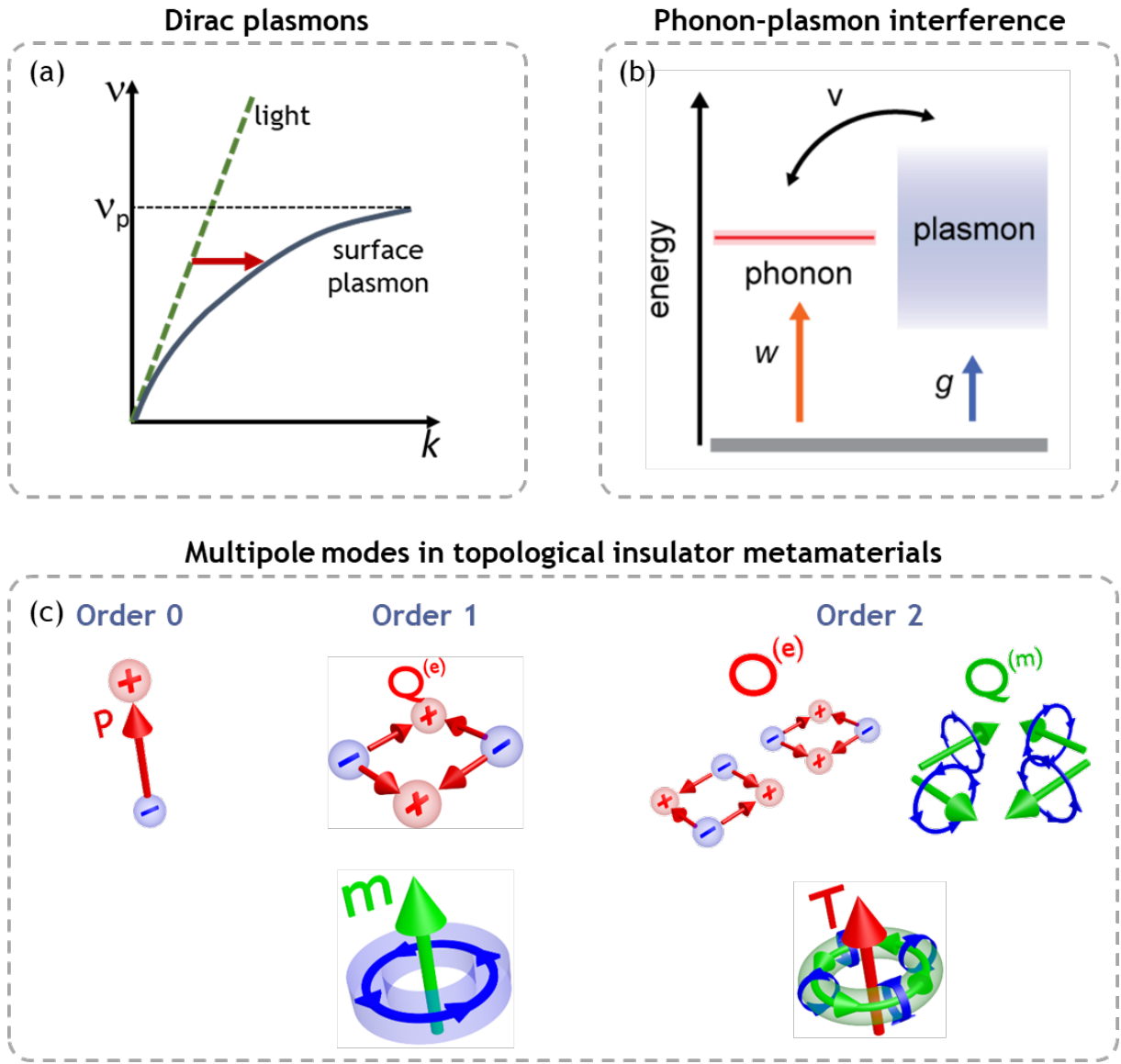

*Figure 16. Interaction of light with topological insulators:* *(a) Schematic of the dispersion plot for a surface plasmon (blue) compared to the light line in vacuum (green). The momentum mismatch (red arrow) makes it necessary to use schemes such as grating structures to excite surface plasmons from carriers in the topological insulator surface or bulk. (b) Scheme of interactions between bulk phonon and plasmons resulting in Fano interference. A discrete energy level of the optical phonon and quasi-resonant plasmon are shown; here, w (g) is the parameter representing coupling between incident photons and the phonon (plasmon) while v determines the Fano interference between the plasmon and phonon. (c) Schematic illustration of complex modes induced in the topological insulator material because of strong infrared polarizability. Figures (b) and (c) are reproduced with permission from In et al.,[162] and Krishnamoorthy et al.,[17] respectively.*

### 4.1. Tunable metamaterials at optical frequencies

We will first discuss metamaterials with frequency tunable response across the optical frequency spectrum from the infrared frequencies to the near-visible, the tunable response arising as a result of change in structural parameters.

### 4.1.1. Metamaterials for the infrared

As seen in Section 2, the far infrared part of the spectrum is characterized by strong plasmonic response from the 2D massless surface state Dirac fermions/3D massive bulk fermions, as well as bulk phonons. Suitably designed metamaterial structures can be employed to elicit responses from both these excitations. The first experimental evidence of 2D plasmonics in a topological insulator ($Bi_2Se_3$) was shown by Di Pietro et. al in a metamaterial composed of micro-ribbon/grating arrays (Figure 17a).[100] Thin micro-ribbon arrays of different widths W and periods 2W were used to select suitable values of the plasmon wavevector k. It was shown that the plasmon dispersion curve agrees quantitatively with that predicted for photons coupled to the 2D Dirac carriers and differs markedly from that due to 3D carriers (Figure 17b). In addition, the resonant extinction spectra of the micro-ribbon arrays were found to exhibit a Fano-like resonance feature due to the interaction between the narrow phonon modes and the broader plasmon mode. In another work, strong plasmon-phonon hybridization via Fano interference was shown in a micro-ring metamaterial structure in $Bi_2Se_3$. The metamaterial exhibited bonding and antibonding plasmon modes whose frequencies were tuned by varying the micro-ring diameter.[186] A similar micro-ribbon array metamaterial design was subsequently employed to demonstrate quantum phase transition from the topological to a trivial insulating phase[188] in $(Bi_{1-x}In_x)_2Se_3$ topological insulator. The material was tuned from topological to trivial insulating state by In doping onto the Bi-site which resulted in a marked change in the linewidth of the plasmon across the topological transition, a strong indication that plasmons are protected in the topological phase.

Coupling of Dirac plasmons on the top and bottom surfaces of thin $Bi_2Se_3$ films was shown by Ginley, Law and co-workers.[192] Plasmons were excited by patterning the films into microribbon arrays and the plasmon frequency dependence on both film thickness and stripe width was shown to be in good agreement with theoretical models for 2D Dirac plasmons. The metamaterial was found to have effective mode indices exceeding 200, demonstrating extremely high confinement of light in the THz frequency range. A system of multiple coupled Dirac plasmons were created by growing topological insulator/band insulator heterostructure composed of alternating layers of $Bi_2Se_3$ (topological insulator) and $(Bi_{1-x}In_x)_2Se_3$ (band insulator) and patterning them into stripe arrays.[193] This structure exhibited excitation of plasmon modes that couple to the phonons in the superlattice, resulting in hybrid plasmon-phonon polaritons, as well as an epsilon near-zero mode in the top dielectric material $(Bi_{0.5}In_{0.5})_2Se_3$.

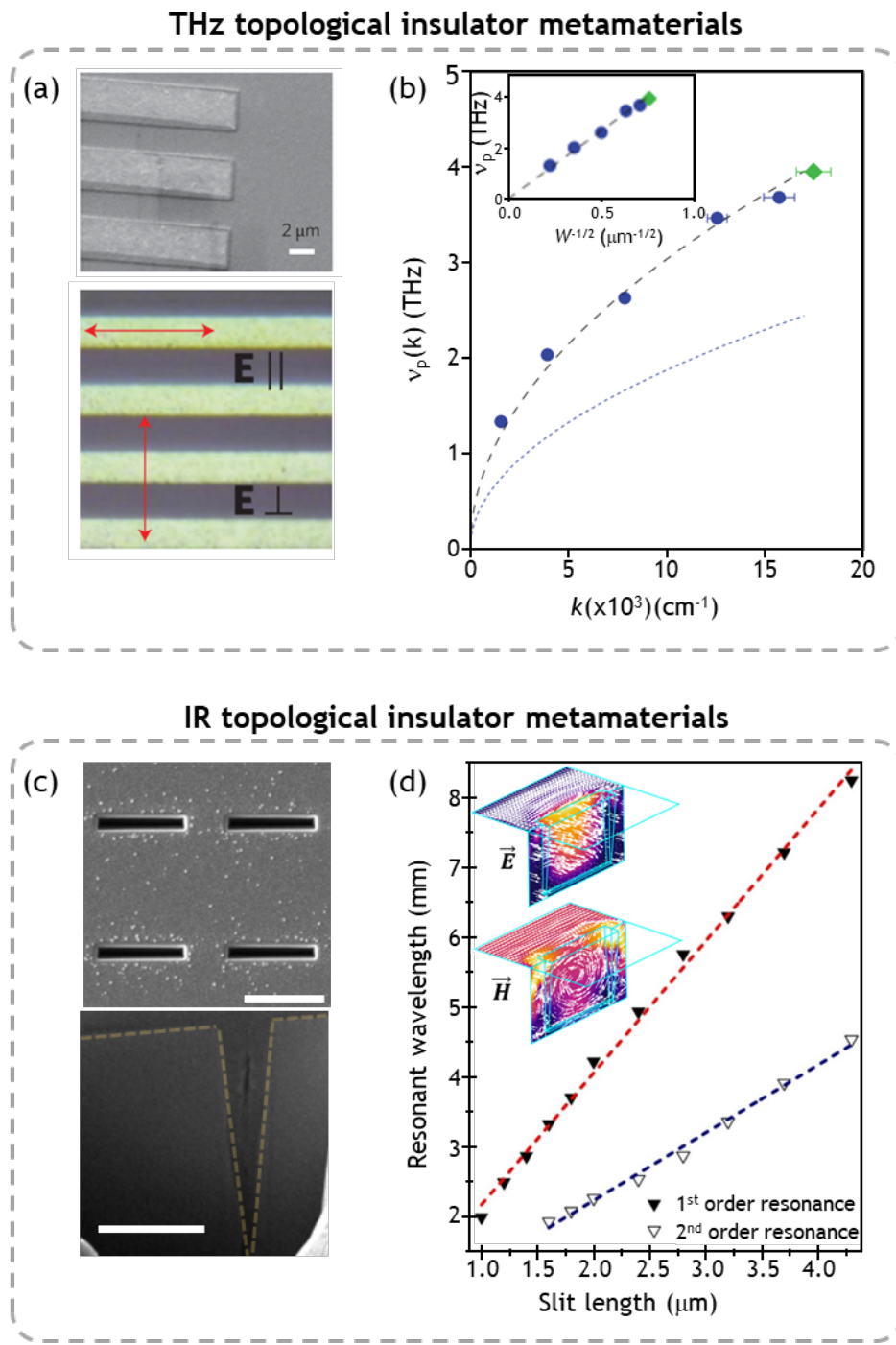

*Figure 17. THz and infrared topological insulator metamaterials:* (a) SEM (top) and optical (images) of micro ribbon arrays patterned in $Bi_2Se_3$ showing also the direction of the incident electric field, either parallel or perpendicular to the ribbons. Dispersion of plasmon modes excited in the micro ribbon arrays with the experimental data (dots) showing agreement with calculated dispersion of 2D Dirac plasmons (dashed black

*curve) and mismatch with dispersion of 3D plasmons (dotted blue curve). (c) SEM images of infrared topological insulator metamaterial composed of nanoslit arrays in $Bi_2Te_3$ image of nanoslit arrays. Scale bar is 4 μm. (d) Resonant wavelength tuning vs slit length for the fundamental and second order resonant modes in the nanoslit arrays. Inset shows the colour maps of electric (**E**) and magnetic fields (**H**) determined by FEM simulations showing the nature of the mode. Figures a, b and c, d reproduced with permission from Di Pietro et al.,[100] and Krishnamoorthy et al.,[17] respectively.*

The long wavelength region is also characterized by a high dielectric polarizability of the bulk as seen in Section 2.2. This is particularly relevant for dielectric metamaterial architectures where a large refractive index translates to a stronger confinement of the optical mode. This was investigated by Krishnamoorthy et al. using nanoslit metamaterial arrays in $Bi_2Te_3$ with resonances spanning the near- and mid-infrared regions (Figures 17c, d).[17] Strong resonant modulation of the incident electromagnetic field in the 2–10 μm region was observed, with analysis of the complex resonant mode structure alluding to the excitation of circular surface currents. This was suggested as a route to couple to the spin-polarized topological surface carriers, thereby providing new opportunities to combine dielectric, plasmonic and magnetic metamaterials in a single platform.

### 4.1.2. Metamaterials for near-visible frequencies

Near visible frequencies, the optical properties of BSTS topological insulator materials are governed by bulk interband transitions showing plasmonic behaviour in the visible region and strongly dielectric behaviour in the near-infrared, as discussed in Section 2.2. There have been several works that employed metamaterial strategies to demonstrate peculiar phenomena in this region. Ou et. al., was the first to demonstrate topological insulator metamaterials composed of nanoslit arrays and grating arrays in $Bi_{1.5}Sb_{0.5}Te_{1.8}Se_{1.2}$ operating up to the ultraviolet region (Figures 18a, b).[187] The metamaterials showed strong absorption resonances with dependence on the geometrical structure of the unit cell as well as the polarization of the incident light. In particular, nanoslit arrays showed strong colours when viewed in reflection mode with incident light polarized perpendicular to the long axis of the nanoslit, while grating arrays showed strong cathodoluminescence peaks. Dubrovkin et. al., showed phase-resolved real-space imaging of localized plasmons in $Bi_{1.5}Sb_{0.5}Te_{1.8}Se_{1.2}$ topological insulator nanostructures by performing scattering-type scanning near-field optical microscopy.[99] Dipolar and higher order surface plasmon modes with well-defined field amplitude and phase profiles were revealed in the visible part of the spectrum in nanostructured topological insulator obtained from focused ion beam milling as well as unstructured nanometer-sized topological insulator flakes (Figures 18c, d). Zhao and co-workers studied a lithography-free metamaterial consisting of $Bi_2Te_3$ nanoplates in solution and demonstrated multiple plasmon modes covering the entire visible range by means of electron energy loss spectroscopy and cathodoluminescence spectroscopy.[159] Different modes were observed in the center and at the edge of the nanoplate, and a dark breathing plasmon mode was excited in the center, interpreted to be the first such instance in a plasmonic material other than noble metals (Figures 18e, f).

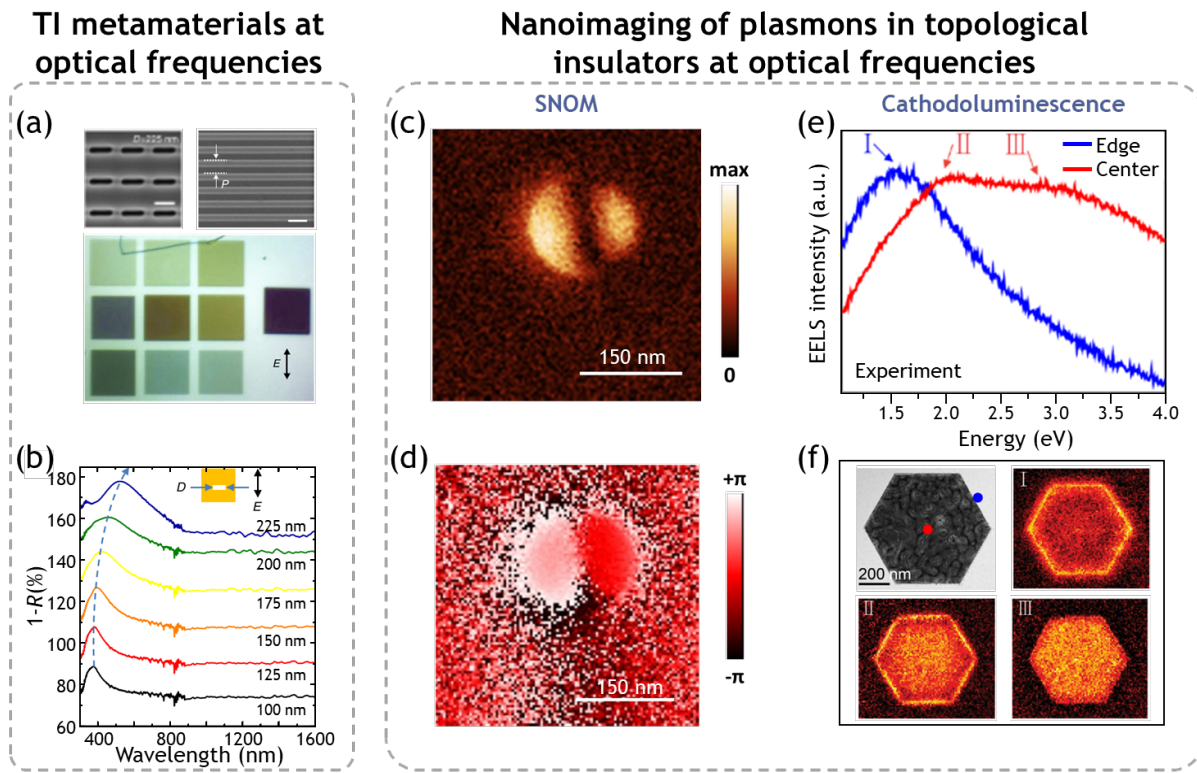

*Figure 18. Topological insulator metamaterials at optical frequencies:* (a) SEM images of nanolsit (top left) and nanograting (top right) arrays fabricated in $Bi_{1.5}Sb_{0.5}Te_{1.8}Se_{1.2}$. Scale bars, 200 nm (left), 500 nm (right). Optical microscopy image (bottom) of nanoslit arrays illuminated with light polarized perpendicular to the slit's long axis showing plasmonic colour. (b) Absorption spectra, 1-R, of the fabricated nano-slit arrays with lengths D from 100 to 225nm for light polarized perpendicular to the slit's long axis. (c) and (d) Near-field amplitude and phase images, respectively, of $Bi_{1.5}Sb_{0.5}Te_{1.8}Se_{1.2}$ nanodisk measured at 633nm excitation wavelength. (e) Electron energy loss spectrum (EELS) of hexagonal $Bi_2Te_3$ nanoplate when the electron beam is positioned at the middle of the edge (blue line) and center (red line) of the nanoplate, corresponding to the blue and red dots in the picture of top left of panel f. (f) EELS map measured experimentally at the three positions indicated in panel e. Figures a, b reproduced with permission from Ou et al.[187] Figures c, d reproduced with permission from Dubrovkin et al.[99] Figures e, f reproduced with permission from Zhao et al.[159]

Optical range metamaterials in topological insulators have found a range of applications. Strong plasmonic absorption resonances were shown by Yue et al. in focused ion beam-milled $Bi_{1.5}Sb_{0.5}Te_{1.8}Se_{1.2}$ nanocone metamaterials with a 15% enhancement in absorption in the ultraviolet and visible regions for potential applications in photovoltaics.[194] Furthermore, angular-momentum nanometrology was demonstrated through the spatial displacement engineering of plasmonic angular momentum modes in a CMOS-compatible plasmonic topological insulator material by Yue et al.[195] Asymmetric circular nanogroove design in $Sb_2Te_3$ topological insulator film was employed to couple optical beams with different angular momentum modes into linearly displaced plasmonic fields and implement angular momentum nanometrology with a low crosstalk of less than −20 dB, opening opportunities for on-chip manipulation of optical angular momentum. In another demonstration, the unique properties of topological insulator materials were exploited to create extremely thin holographic films.[196] Nanometric topological insulator thin films were shown to act as an intrinsic optical resonant cavity due to the unequal refractive indices in their metallic surfaces and bulk, leading to enhancement of phase shifts and realization of holographic imaging. Aigner and co-workers recently predicted that the superior plasmonic figure of merit of $Bi_2Te_3$ topological insulator can be utilized to create gap plasmon metasurfaces and designed a far-field beam-steering metasurface for the visible spectrum, yielding a cross-polarization efficiency of 34% at 500 nm while suppressing the co-polarization to 0.08%.[197]

### 4.2. Active and reconfigurable metamaterials

As discussed in Section 2.2, modulation of the optical response of topological materials with spin-momentum locking can be exploited for the realization of new types of active and reconfigurable metamaterials. Some specific modulation mechanisms include: (i) optical carrier injection by above-gap optical frequency excitation which creates photoexcited bulk and surface carriers or below-gap THz excitation which creates photoexcited surface carriers; (ii) optical nonlinearities of the topological surface states that can be induced by symmetry breaking at the crystal surface or by selective THz excitation below-gap; (iii) electro-optic modulation of the surface response by application of external biases that vary the Fermi level; (iv) magneto-optic modulation of plasmon and cyclotron responses by application of external magnetic fields; and (v) directional photocurrents induced by excitation with circularly polarized light. The modulation of the optical response of topological materials induced by these mechanisms can be further amplified and controlled by structuring them with appropriate metamaterial designs.

In the following, we survey recent demonstrations of active and reconfigurable metamaterials realized in topological insulators.

### 4.2.1. Ultrafast pulse induced modulation of metamaterial resonances

Thin, low-power, ultrafast systems are in demand for realizing ultrafast modulators particularly for optical communication. Topological materials, with their unique blend of bulk and surface features facilitate the development of such systems. In et. al., proposed a methodology for controlling electron−phonon interaction in lithographically engineered Dirac surface plasmons in the $Bi_2Se_3$ topological insulator.[190] Optical pump THz probe time-resolved ultrafast measurements in such $Bi_2Se_3$ slit array metamaterials revealed a shift in the plasmon resonance frequency induced by the pump pulse (Figure 19a). Time-dependent analysis of the mechanisms involved including Dirac plasmon behaviour, phonon broadening, and phonon stiffening revealed a transition between two distinct dynamic regimes as the Dirac plasmon resonance of the metamaterial moves across the topological insulator phonon resonance, thereby realizing dynamic control of the interaction between Dirac carriers and phonons. Relaxation dynamics of the 2D electron gas associated with the Dirac plasmon showed decay times ~ few ps implying that ultrafast modulation can be achieved. In a prior work, the same group demonstrated extremely high modulation depth of ~2400 % in a $Bi_2Se_3$ microribbon THz metamaterial as a function of the delay between the optical pump and the THz probe beams, reflecting the dynamics of photoexcited Dirac carrier population responsible for the resonant plasmonic mode.[198] Grating, asymmetric split ring resonator THz metamaterials with strong dipole and Fano resonances were also shown recently in the topological semimetal $Cd_3As_2$ by Dai et al.[199] Experimental studies revealed low power, ultrafast modulation (switching time ~15 ps) and long momentum scattering rate (~150 fs). In essence, combining metamaterial strategies with unique characteristics of topological materials open routes to realize ultrafast optical modulators.

### 4.2.2. Topological surface state nonlinearity in metamaterials

Nonlinear phenomena play a fundamental role in modern photonics, enabling optical functionalities like ultrashort pulse generation and shaping, sum- and difference-frequency processes and ultrafast switching. Di Pietro et al. demonstrated that the strong nonlinear optical response of topological surface states manifests on the plasmon itself, which can be further enhanced in artificially structured metamaterials. With increasing field, the bare plasmon frequency of a 20 μm-ribbon film was found to shift from 1.5 THz at 0.1 kV/cm (the linear regime value) to 0.9 THz at the highest field (~1 MV/cm), i.e., 60% of its linear value (Figure 19b). This behaviour was attributed to the low heat capacity associated with the massless Dirac electrons by a thermodynamic model. Here, the temperature of the electron bath undergoes a strong enhancement upon intense THz excitation, resulting in a nonlinear response dominated by incoherent thermal effects.[200] This plasmonic softening behaviour contrasts with optical pump-THz probe results wherein the plasmon hardens due to the effective increase of the Dirac carrier population when the pump energy is above the optical gap.[198] Topological insulator metamaterial systems thus function as a unique platform to generate reconfigurable nonlinear optical phenomena, particularly in the THz region where there is a lack of suitable nonlinear material systems.

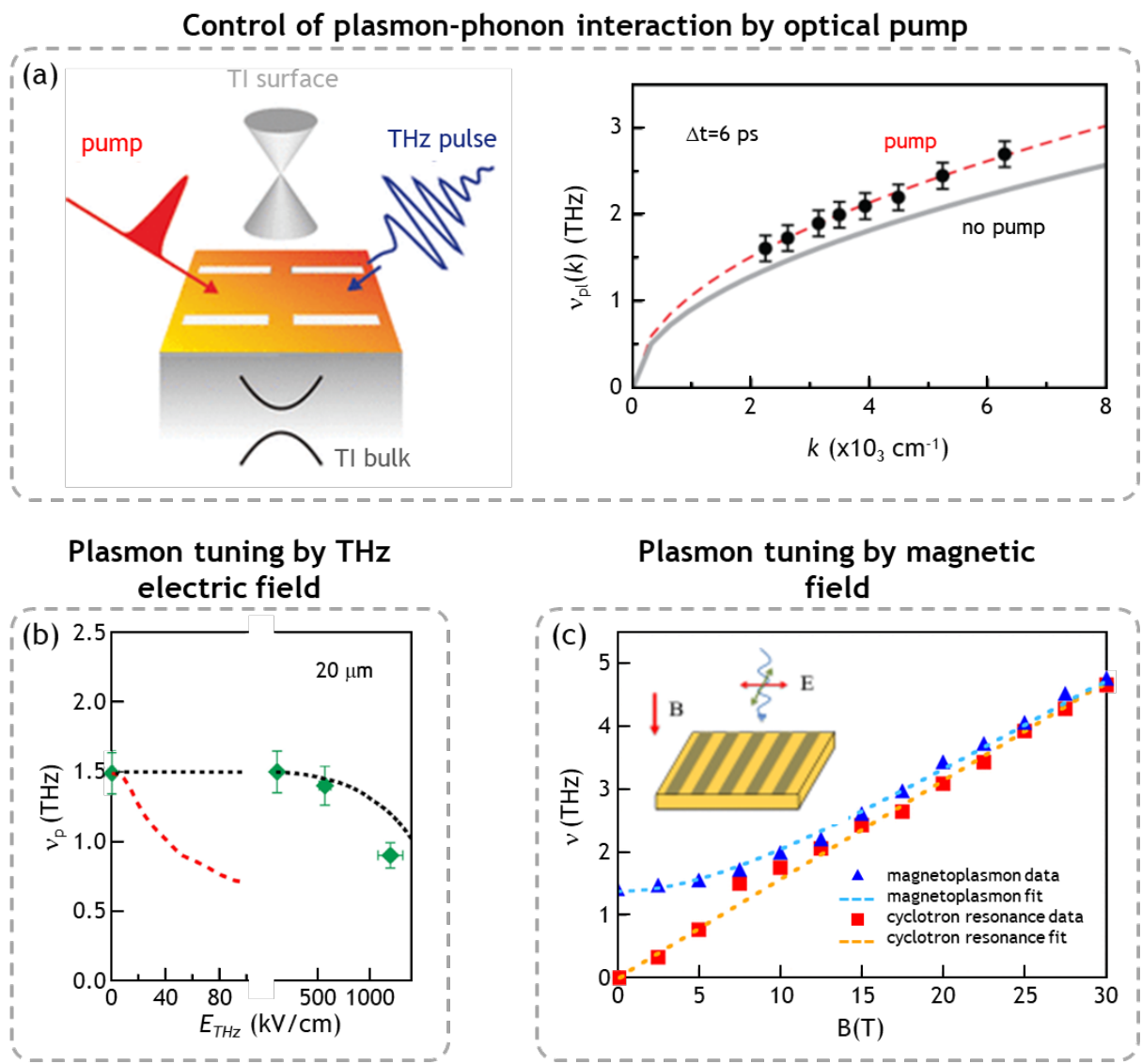

*Figure 19. Active topological insulator metamaterials: (a) Experimental schematic of an optical pump-THz probe setup for controlling phonon-plasmon interactions in a Bi$_2$Se$_3$ slit metamaterial.[190] Plasmon dispersion after optical excitation, at delay time Δt = 6 ps (black circles) together with a fit (red dashed curve). The solid grey curve shows the dispersion of the plasmon dispersion measured without the pump pulse. (b) Experimental plasmon frequency v$_p$ as a function of the THz electric field for micro ribbon structure of ribbon width 20 μm showing plasmon softening as the intensity of the electric field is increased.[189] The green dots are the experimental data while the black and red dashed curves are calculated behaviour based on a thermodynamic model and*

*Boltzmann equation, respectively. (c) Observation of magnetoplasmons in ribbon arrays in $Bi_2Se_3$[191] – experimental (blue triangles) and fitted behaviour (blue curve) of magnetoplasmon frequencies as function of magnetic field B. Experimental (red squares) and fitted behaviour (orange curve) of cyclotron frequencies as a function of B field. Figure a, b, and c are reproduced with permission from Di Pietro et al.,[189] Autore et al.,[191] and In et al.,[190] respectively.*

### 4.2.3. Magnetic field induced modulation of metamaterial resonances

Surface plasmon modes whose properties can be controlled by an external magnetic field are highly desirable for optospintronic applications. Autore et al. investigated the behaviour of Dirac plasmon modes excited in $Bi_2Se_3$ ribbon arrays under a strong magnetic field measuring both the plasmon and cyclotron responses (Figure 19c).[191] At low magnetic fields, the cyclotron resonance is still well separated in energy from the magnetoplasmon mode, while at high magnetic fields both excitations asymptotically converge at the same energy, thus demonstrating the ability to magnetically control plasmonic excitations. This could open the way towards realization of plasmon-controlled magneto-optic devices.

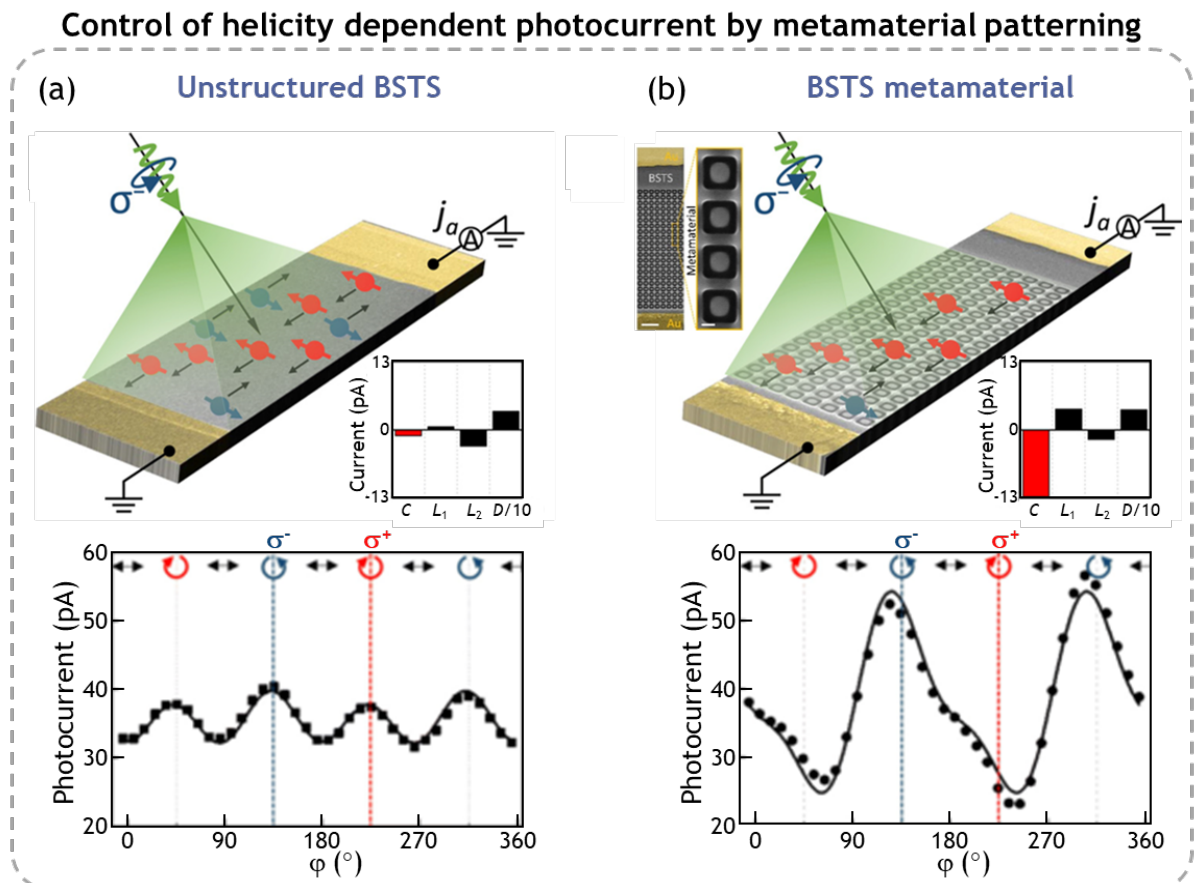

*Figure 20. Photocurrents in topological insulators:  a) Helicity dependent photocurrent (HDPC) in a $Bi_{1.5}Sb_{0.5}Te_{1.8}Se_{1.2}$ topological insulator at room temperature: CPGE is masked by other contribution.[201] b) Artificial nanostructuring of BSTS topological insulator with an achiral metamaterial design dramatically enhances the circular photogalvanic effect (CPGE).[201] Figures reproduced with permission from Sun et al.[201]*

### 4.2.4. Light helicity-dependent photocurrents in resonant metamaterials

Nanophotonic strategies can be used to create and enhance chiral optical response in designer metamaterials. Sun et al. reported the first experimental demonstration of a topological insulator metamaterial in which the intrinsic circular photogalvanic effect (CPGE) was enhanced by nanostructuring the crystal (Figure 20a).[201] Tight confinement of electromagnetic fields provided by resonant achiral metamaterial carved into $Bi_{1.5}Sb_{0.5}Te_{1.8}Se_{1.2}$ topological insulator crystals effectively increased selective photoexcitation of spin-polarized surface states under circular polarized light illumination, resulting in a giant enhancement of the CPGE response (Figure 20b). Additionally, Sun et al. have shown that mirror-symmetric planar chiral metamaterials can be used to control the ratio between spin photocurrents flowing in opposite directions, thanks to the circular dichroism induced by the metamaterial designs.[16] The implementation of photonic metamaterials to enhance, isolate and manipulate the spin-polarized surface states of topological insulators provides new ways to control light-matter interaction in topological materials and paves the way towards realization of chiral photodetectors with tailored spectral and polarization response.

### 5. Conclusions and perspectives

Despite its potential, the field of topological insulator metamaterials is just at its infancy. Further advancement of the field requires improvement in materials systems and quality, film characteristics as well as novel metamaterial designs. For example, new materials with higher surface to bulk response and novel optoelectronic properties could provide enhanced metamaterial responses. In addition, easier means to grow or procure large area films with controlled thicknesses can significantly help in expanding the gamut of metamaterial designs being employed. Another point of concern is the sensitivity of the topological properties to exposure to ambient conditions or other chemicals during nanofabrication/processing steps; suitable methodologies need to be employed to mitigate this issue, for example using appropriate protective layers,[19] or recovering surface states through thermal annealing. In essence, improved reliability, reproducibility in materials synthesis and fabrication for optical applications as well as extension of strong surface state optical response up to the visible and near infrared frequencies would open exciting opportunities in modern nanophotonics.

To conclude, we would now like to provide our perspectives on interesting research directions for this field of topological insulator metamaterials.

*Non-Hermitian photonics*

Conventionally, optical loss is considered to be a nuisance that limits the applicability of photonic devices. The notion of non-Hermiticity looks to utilize these losses to introduce novel functionalities in photonic devices and relies on the fact that in non-Hermitian systems respecting parity time (PT) symmetry, it is possible to have real eigen values.[202,203] Several topological material systems, for example, chalcogenide topological insulators are characterized by interband absorption and accompanying losses in parts of the optical frequencies. Combining such materials with suitable gain media and nanostructuring strategies, one may be able to realize non-Hermitian photonic systems characterized by PT- and anti-PT-symmetries to enable various functionalities such as optical power splitting and modulation, chirality and topological effects.

*Hybrid electronic-photonic topological insulator systems*

The emerging field of topological photonics concerns the design of photonic structures that mimic topological electronic systems for localization and transport of photons via edge or surface states.[29-31] Electronic topological insulators such as Bi/Sb trichalcogenides possess extremely high index of refraction, which makes them suitable for the realization of dielectric topological photonic structures. This provides the opportunity to couple topologically protected plasmons in the electronic topological insulator with the topologically protected optical modes of the photonic system, opening new avenues for integrated plasmonic circuits. Use of alternative platforms such as Weyl semimetals may also allow devising complex systems based on their exotic 3D topological plasmons.

*Chiral superconducting single photon detectors*

Detectors based on superconducting materials have recently established themselves as the best choice for single photon applications owing to their high sensitivity, detection efficiency as well as fast response. Topological materials with strong photocurrent responses enhanced by metamaterials are excellent candidates for chiral light detection.[204] A topological superconductor can potentially combine the superior features of the typical superconducting single photon detectors with the chiral sensitivity arising from the topological bands.

*Hyperbolic and chiral topological plasmonics*

Materials characterized by a hyperbolic optical isofrequency contour enable strong light-matter interaction with applications in single photon emission, sub-diffraction imaging and biosensing.[85,205-207] Originally shown in artificially engineered structures over broadband frequencies, these features have also been shown to occur in naturally occurring materials over a limited range of frequencies. Recently, Weyl semimetals have been added to the class of natural hyperbolic materials. Theory has predicted that the collective dynamics of electrons on both the open-segment and closed Fermi surfaces in Weyl semimetals with broken time reversal symmetry leads to the creation of Fermi arc plasmons with hyperbolic dispersion and chiral nature, that could allow for tight focusing of collimated, nonreciprocal surface plasmon waves with frequency-dependent directionality.[132] Such hyperbolic plasmons have been demonstrated in $WTe_2$ and ZrSiSe very recently.[138,139] Furthermore, Weyl semimetals in the presence of a magnetic field are predicted to support two types of highly confined 3D topological plasmons with anisotropic dispersions, from linear to parabolic or even hyperbolic bands.[134] Hybridizing such materials with suitable metamaterial structures could provide additional degrees of freedom to control and enhance light-matter interaction.

After reviewing the growth, fabrication and electromagnetic properties of various topological material systems, we have presented artificial structuring as a practical means to harvest their exotic topological properties in the photonic domain. General features of the surface states of topological insulators such as spin-polarized texture and protection from scattering have been effectively combined with metamaterial strategies to realize hybridized phonon-plasmon modes, enhanced photocurrents, nonlinear and magneto-optic effects. Topological semimetals could further enhance these effects by providing additional degrees of freedom such as 3D topological plasmons together with hyperbolic isofrequency contours and large chirality. Moreover, topological superconductors now at their early stages in the context of photonics, have great potential for THz applications. All-in-all, we envision that further advances in topological materials discovery and artificial structuring, along with the development of more complex metamaterial designs could lead to the uncovering of new physical phenomena and realization of functional metamaterial devices hybridizing photons with spin-polarized Dirac carriers at the nanoscale.

**Author's short biography**

**Harish N. S. Krishnamoorthy** is a Senior Research Fellow at the Centre for Disruptive Photonic Technologies at Nanyang Technological University, Singapore. He received his M.Sc. degree in Photonics (2007) from the Cochin University of Science and Technology (India) and M. Phil (2012), Ph.D. (2014) degrees in Physics from the City University of New York (USA). His research interests include quantum and phase change materials, light-matter interaction engineering, as well as nano and micro photonics.

**Alexander M. Dubrovkin** is a Senior Research Fellow at the Centre for Disruptive Photonic Technologies at Nanyang Technological University (Singapore). His research interests include 2D materials, nanophotonics and nanoscale optoelectronics, topological physics and light-matter interaction at the atomic scale. He received M.S. (2007) and Ph.D. (2011) degrees from Lomonosov Moscow State University (Russia).

**Giorgio Adamo** is a Giorgio Adamo is a Principal Research Fellow at Nanyang Technological University (NTU), Singapore. He received his Ph.D. degree in 2011 from the University of Southampton (UK) and then moved to Singapore where he was key in establishing the Centre for Disruptive Photonic Technologies (CDPT) at NTU. The topics of his research include light- and free electron- interaction with nanostructured matter, metamaterials and metasurfaces, quantum and topological nanophotonics.

**Cesare Soci** is an Associate Professor of Physics and Electrical and Electronic Engineering at Nanyang Technological University. He received Laurea (2000) and Ph.D. (2005) degrees in Applied Physics from the University of Pavia. He was a postdoctoral researcher at the Center for Polymers and Organic Solids of the University of California, Santa Barbara (2005-2006), and at the Electrical and Computer Engineering Department of the University of California, San Diego (2006-2009). In 2009, he moved to Singapore where he established the Optical Spectroscopy of Nanomaterials laboratory and co-founded the Centre for Disruptive Photonic Technologies. His research focuses on optical materials for nanophotonics. He is a Fellow of the IPS, the OSA and the SPIE.

**Acknowledgments**

This research was supported by the Singapore Ministry of Education (Program MOE2016-T3-1-006), and the Quantum Engineering Programmes of the Singapore National Research Foundation (QEP-P1 and NRF2021-QEP2-01-P01).

**List of abbreviations**

1D – One dimensional

2D – Two dimensional

3D – Three dimensional

AFM – Atomic Force Microscopy

BSTS – $Bi_xSb_{2-x}Te_ySe_{3-y}$

BZ – Brillouin zone

CMOS – Complementary metal–oxide–semiconductor

CPGE – Circular photogalvanic effect

DFT – Density Functional Theory

EELS – Electron Energy Loss Spectroscopy

HDPC – Helicity dependent photocurrent

IR – Infrared

MBE – Molecular beam epitaxy

PT – Parity time

QH – Quantum Hall

QSH – Quantum Spin Hall

SEM – Scanning Electron Microscopy

SHG – Second Harmonic Generation

SOC – Spin-orbit coupling

THz - Terahertz

VLS – Vapor liquid solid

VS – Vapor solid

**Table of contents (TOC) graphic**

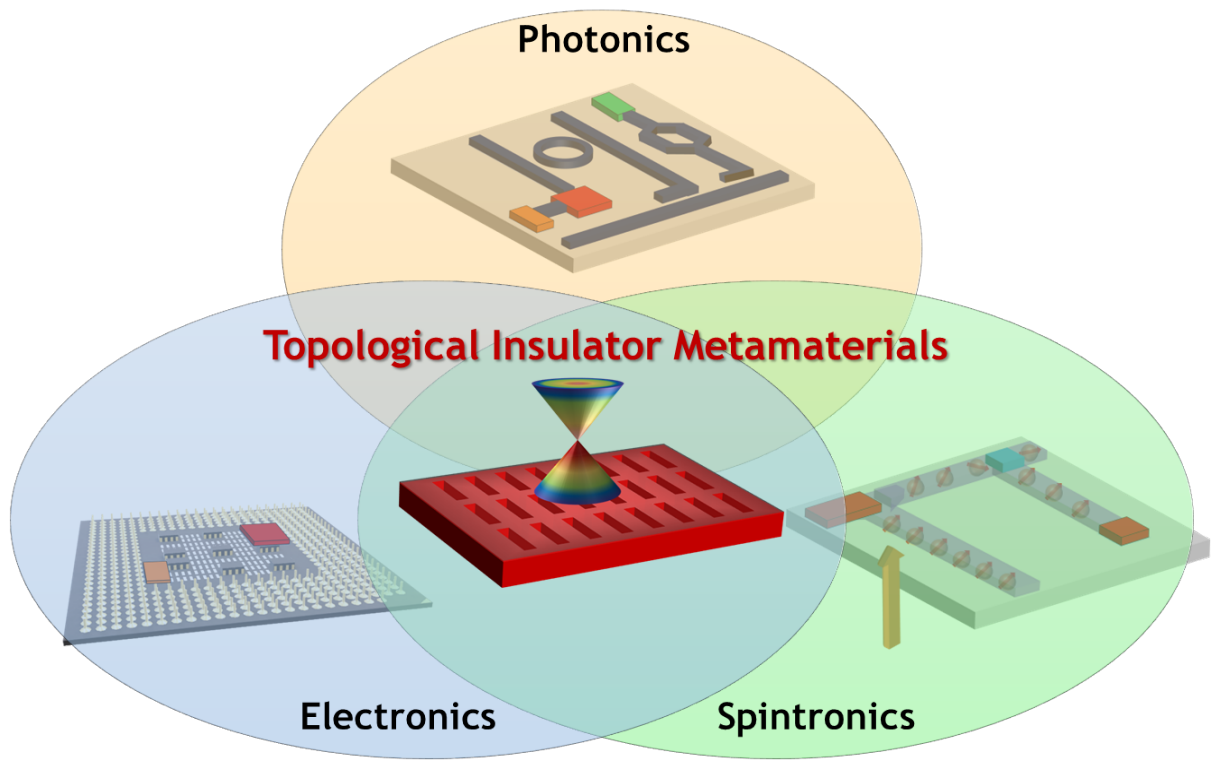